\documentclass[final,1p,times]{elsarticle}
\usepackage[table,xcdraw]{xcolor}
\usepackage{graphicx}
\usepackage{hyperref}
\usepackage{amsmath}
\usepackage{amssymb}
\usepackage{amsfonts}
\usepackage{multirow}
\usepackage{booktabs}
\usepackage{array}
\usepackage{multirow}
\usepackage[flushleft]{threeparttable}
\usepackage{tikz,float}
\usepackage{xcolor}
\usetikzlibrary{positioning, shapes.geometric}
\usepackage{enumitem}
\usepackage{array}
\usepackage{natbib}
\usepackage{graphicx}
\usepackage{lscape}
\usepackage{adjustbox}
\usepackage{tabularx}
\usepackage{svg}
\usepackage{cleveref}
\usepackage{comment}
\usepackage{tikz}
\usepackage{pdfrender}
\usepackage{rotating}
\usepackage{color,soul}
\usepackage{multicol}

\renewcommand\hl[1]{#1}

\journal{Internet of Things Journal}

\begin{document}

\newcommand*{\boldcheckmark}{%
  \textpdfrender{
    TextRenderingMode=FillStroke,
    LineWidth=.5pt, 
  }{\checkmark}
}

\begin{frontmatter}


\title{AnoML-IoT: An End to End Re-configurable Multi-protocol Anomaly Detection Pipeline for Internet of Things}



 \author[CU]{Hakan Kayan}
 \author[CU]{Yasar Majib}
 \author[CU]{Wael Alsafery}
 \author[FR]{Mahmoud Barhamgi}
 \author[CU]{Charith Perera}
       
 \address[CU]{Cardiff University, UK}
 \address[FR]{Claude Bernard Lyon 1 University, France}

%
%
%
%
\begin{abstract}
The rapid development in ubiquitous computing has enabled the use of microcontrollers as edge devices. These devices are used to develop truly distributed IoT-based mechanisms where machine learning (ML) models are utilized. However, integrating ML models to edge devices requires an understanding of various software tools such as programming languages and domain-specific knowledge. Anomaly detection is one of the domains where a high level of expertise is required to achieve promising results. In this work, we present AnoML which is an end-to-end data science pipeline that allows the integration of multiple wireless communication protocols, anomaly detection algorithms, deployment to the edge, fog, and cloud platforms with minimal user interaction. We facilitate the development of IoT anomaly detection mechanisms by reducing the barriers that are formed due to the heterogeneity of an IoT environment. The proposed pipeline supports four main phases: (i) data ingestion, (ii) model training, (iii) model deployment, (iv) inference and maintaining. We evaluate the pipeline with two anomaly detection datasets while comparing the efficiency of several machine learning algorithms within different nodes. We also provide the \href{https://gitlab.com/IOTGarage/anoml-iot-analytics}{\color{blue}source code} of the developed tools which are the main components of the pipeline.
\end{abstract}

\end{frontmatter}
\section{Introduction}
\label{Sec:Introduction}
Edge AI which is critical for resource-constrained environments that operates in the Internet of Things (IoT) domain where intelligent tasks are performed has started to become a hot topic with the arrival of Industry 4.0 \cite{lasi2014industry}. It manages the interaction with the physical world that is provided by sensors and actuators. Management of such an environment requires series of tasks (e.g., data collection, anomaly detection) that are operated by microcontrollers running ML models. Data-related professions (e.g., data scientists, ML engineers) define rules/ranges and search for the best practices to increase the operability of edge mechanisms in their relevant scientific disciplines. Finding hidden information from big data can enhance the quality of living but it is not a straightforward task \cite{Mohammadi2018}.

While for data scientists, being an expert in edge-related infrastructures (e.g., programming languages, microcontrollers, sensors) is not expected, they should be able to utilize data science pipelines which are executable workflows of data-related tasks that automate the desired process. Thus, we developed a reconfigurable data science pipeline based on an IoT sensing infrastructure that utilizes open-source software to facilitate developing an interconnected anomaly detection system that runs on edge, fog, and cloud platforms. We define the edge as the platform where the first interaction between the cyber and physical world happens. Hence, microcontrollers (e.g., Raspberry Pi Pico) that gather physical data are edge devices. We define fog as the platform where several edge devices can be supervised. Hence, single-board computers (e.g., Raspberry Pi 4B) are fog devices that might act as edge devices as well. Cloud is the platform where real-world data gathered by the edge and fog devices are progressed. We implemented our system based on an example use case scenario to describe how to proposed system works while providing some results.

The contributions of this paper are as follows:

\begin{itemize}

\item We provide reconfigurable IoT sensing infrastructure that consists of two main open-source components: (i) The Edge to Cloud Code Generator (EECG) that generates ready-to-deploy codes to enable data circulation from edge to fog. (ii) The Node-RED package is hosted on Node-RED servers that enables accessing and processing to the edge data from anywhere that has access to Node-RED servers while offering visualization via the graphical user interface (GUI). We also provide one Python library and executable shell script that facilitate data training and inference phases.

\item We propose a data science pipeline that interconnects edge, fog and cloud devices/services to provide end-to-end anomaly detection system development. The pipeline contains four main stages: (i) the data collection which is provided by the components mentioned at the first contribution point, (ii) the anomaly detection model training, (iii) model deployment to the edge, fog, and cloud, (iv) inference, and maintaining the model based on the new data. We demonstrate how the proposed tools are utilized during these stages.

\item \hl{We provide a dataset that is generated via the utilization of proposed tool. We analyze the performance of Convolutional Neural Network (CNN)} \cite{jiang2020convolutional}, \hl{Recurrent Neural Network (RNN)} \cite{goh2017anomaly}, \hl{Isolation Forest} \cite{liu2008isolation} \hl{and One-class Support Vector Machines (OC-SVM)} \cite{ma2003time} \hl{on the proposed dataset} \cite{dataset} \hl{and the WADI dataset} \cite{ahmed2017wadi}. \hl{We also evaluate them according to the platform (edge, for or cloud) where the anomaly detection model is deployed via the utilization of proposed pipeline. In the edge, we only evaluated CNN due to lack of application programming interface (API).}

\end{itemize}

\textbf{Structure of the Paper:}
This section provides a high-level understanding of what we proposed. We outlined the previous commercial and academic works in section \ref{Sec:RelatedWork}. Section \ref{section:Architecture} contains the architecture of an IoT anomaly detection pipeline infrastructure. Section \ref{section:dataCirculation} presents details about how the data circulated and progressed within the AnoML-IoT pipeline. We demonstrate our evaluations and results in section \ref{Sec:evaluation2}. Then, we discuss about the results in the section \ref{Sec:lessonslearnt} and finally provide our conclusions in section \ref{Sec:Conclusion}.

\section{Related Work}
\label{Sec:RelatedWork}

In this section, we introduce the data science pipelines that are offered either by academia or commercial entities, and anomaly detection techniques in time-series sensor data. We also analyze the capabilities of open-source platforms that facilitates data circulation.

\subsection{Unsupervised Anomaly Detection in Time Series Sensor Data}
\label{Sec:anomalyDetection}

Anomaly detection is one of the fundamental fields that utilize the machine learning (ML) model as the main component. There is extensive research being done in this field \cite{Liu2020, Chalapathy2019, Blazquez-Garcia2020}. There are three types of anomalies: (i) point anomalies, (ii) contextual anomalies, (iii) collective anomalies. If the anomalies are contextual where the context is time, the time series anomaly detection models are applied. For example, in an environment where the weather temperature decreases at night if the temperature value generated by the sensor acts otherwise, there is a contextual anomaly. While point anomalies are easier to detect, contextual and collective anomaly detection requires additional tasks to identify the normal behavior of the system.

The nature of the input data is the core element that determines the efficiency of the ML model. The features of the data may depend on several complementary terms such as labels, context, and domain. For example, if the input data do not contain any labels that define normality, unsupervised algorithms \cite{song2013toward} are applied, if the data is related to a certain context, context-aware \cite{zhu2012context} methods are selected, if the environment is industrial, because of the importance of detection time, faster models with reduced complexity \cite{marti2015anomaly} are preferred.

In an interconnected domain such as IoT, cyber-physical systems \cite{jazdi2014cyber} are utilized to supervise the environment. These systems observe the behavioral changes (e.g., change in the temperature or movement) in surroundings through modules that manage sensors \cite{raghavendra2006wireless}. They can also act as controllers if they contain actuators. In such environments, the anomaly might occur either by independent or dependent events. If the events are independent, univariate analysis \cite{pang2017anomaly} is applied. For example, the behavioral changes in temperature, loudness, light density, and humidity can be detected via the related data only, hence require univariate analysis. However, changes in the angular momentum or acceleration are measured by sensors (e.g., accelerometer, gyroscope) that generate data per dimension. Hence, the relation between the data points should also be analyzed to detect anomalies. Then, the multivariate analysis \cite{su2019robust} is applied.

\begin{figure}[!t]
  \centering
  \includegraphics[scale = 1.00]{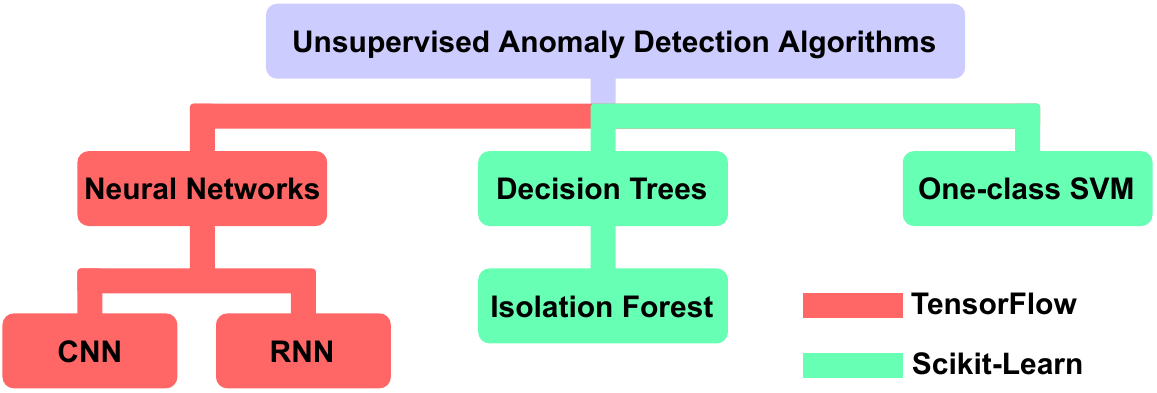}
  \caption{\hl{Illustrates the utilized algorithms. TensorFlow also has an API} \cite{TF_decision} \hl{for decision trees, but due to having better documentation we prefer using scikit-learn for implementing Isolation Forest.}}
  \label{fig:taxonomy}
\end{figure}

While academia keeps offering new anomaly detection algorithms \cite{Teng2010, Wu2017}, most of the time these are based on the fundamental ones \cite{chandola2009anomaly}. Hence, for this work, we selected the following algorithms as they are the most common ones that are utilized for the unsupervised anomaly detection in time series data and accessible via common ML programming libraries/frameworks (e.g., scikit-learn \cite{pedregosa2011scikit}, TensorFlow \cite{abadi2016tensorflow}) : (i) convolutional neural networks (CNN) \cite{munir2018deepant, wen2019time}, (ii) recurrent neural networks (RNN) \cite{li2019mad}, (iii) isolation forest (IF) \cite{liu2008isolation}, and (iv) one-class support vector machines (OC-SVM) \cite{ma2003time}. Figure \ref{fig:taxonomy} demonstrates the used algorithms in this study.

\subsection{Machine Learning Platforms and Data Science Pipelines}
\label{sec:relatedPipelines}

\textbf{Machine learning platforms.}\hl{After showing promising results in a variety of tasks including speech recognition, image processing, anomaly detection, and medical diagnosis, ML has taken an interest in both academic and commercial entities, hence resulting in the creation of many open-source and proprietary ML platforms and pipelines. ML models can be generated via hand-coding, code generators, or interpreters. \textit{Hand coding}. There are many machine learning libraries available} \cite{pedregosa2011scikit, abadi2016tensorflow, ketkar2017introduction} \hl{that allows user to create ML models or to deploy and evaluate ML algorithms. A person with the profession might prefer hand-coding as it offers high customization, allows the development and employment of novel algorithms, and is easy to maintain. However, hand-coding might be very resource-consuming, thus most of the time it is done by a group of programmers. \textit{Code generators}. ML consists of many steps (e.g., data acquisition, data pre-processing, and fitting). Rather than hand-coding all these steps, code generators }\cite{quinnradich_automatic, MATLABML} \hl{might be utilized to facilitate the process. Due to the variety of complicated tasks, most code generators provide a specific code for a specific task. \textit{Interpreters}. One of the main challenges of ML is the portability of the generated model. Interpreters provide portability by generating a model file that can be run on other platforms with minimal coding. TensorFlow} \cite{abadi2016tensorflow} \hl{is the most common one that offers model generation for resource-constrained platforms.}

\hl{\textbf{Data science pipelines.} Raw data are needed to be interpreted to be utilized within data science-related tasks. If the data science pipeline contains all the steps that are required to interpret the data from data gathering to deployment of a machine learning model, it is called end-to-end. These end-to-end pipelines can be either manual where the user provides many inputs and sets parameters each time before a new model is generated or automated where little to no input is taken. Due to a variety of data types, automated pipelines put a certain set of rules (e.g., time format) for their system to accept the input data} \cite{ren2019time}. \hl{These pipelines can also be named according to the performed tasks (e.g., anomaly detection pipeline). Now we introduce pipelines that are presented by either industry or academia.}

\hl{\textit{Azure Machine Learning Pipeline}} \cite{nilspohlmann_create}.\hl{ Microsoft provides an ML pipeline based on running Python scripts on the cloud while automatically handling resource usage. Each step of the pipeline can be independently customized hence offering scalability to the end-user. One of the practical features that Azure Machine Learning Pipeline offers is the automated dependency handling that allows the usage of a variety of hardware and software environments. Microsoft also provides Azure Cognitive Services} \cite{cognitive} \hl{where you can utilize their ML pipeline and Anomaly Detector} \cite{azureAnomaly} \hl{service. They apply Graph Attention Network (GAN)} \cite{zhao2020multivariate}\hl{ for multivariate analysis, apply SR-CNN} \cite{ren2019time} \hl{for the univariate analysis.}

\hl{\textit{Amazon Web Services (AWS) Machine Learning Pipeline}} \cite{aws_build_2021}. \hl{Amazon provides an end-to-end ML pipeline as a service for detecting anomalies in real-time. Inside the pipeline, there are many different services (e.g., database, data formatting) that can be utilized for pipeline tasks. Amazon SageMaker} \cite{liberty2020elastic} \hl{is the main service that provides anomaly detection for both univariate and multivariate data. It allows users to either use a built-in unsupervised anomaly detection algorithm based on Random Cut Forest (RCF)} \cite{guha2016robust} \hl{or use a custom algorithm that can be deployed via a Docker image. Now we introduce the pipelines proposed by the academia.}

\citet{prado2020bonseyes} \hl{propose an end-to-end modular AI pipeline that allows users with less expertise to implement their AI applications such as keyword spotting, image classification, and object detection to systems that contain embedded devices. Their framework relies on Low Power Deep Neural Network (LPDNN) that contains an Inference Engine (LNE) that is compatible with Caffe} \cite{caffe}. \hl{LNE is a code generator that facilitates the deployment to the embedded devices. The authors use FIWARE} \cite{fiware} \hl{for IoT hub integration and Kurento Media server} \cite{fernandez2013kurento} \hl{for media streaming which are required to run live inference. The authors define Raspberry Pi devices as edge and evaluate the efficiency of LPDNN compared TF Lite} \cite{tensorflowLite} \hl{on these devices by running benchmarks that are included in the TF Lite repository.}

\citet{drori2018alphad3m} \hl{propose an automatic ML (AutoML) system that optimizes the ML pipeline according to the given dataset. Their pipeline utilizes LSTM-RNN as a base ML algorithm. Monte Carlo Tree Search (MCTS)} \cite{browne2012survey} \hl{is applied to the predictions generated by the LSTM-RNN to evaluate the performance of the pipeline and decide on the better pipeline. They evaluate the proposed pipeline compared to baseline stochastic gradient descent (SGD)} \cite{bottou2010large} \hl{estimators from scikit-learn} \cite{pedregosa2011scikit}. \hl{They claim their pipeline provides faster run time according to its peers.}

\citet{sutton2018physonline} \hl{propose an open-source ML pipeline that receives physiological data that is used to identify anomalous behaviors as an input in real-time. The authors try to detect Paroxysmal atrial fibrillation (PAF) by applying Probabilistic Symbolic Pattern Recognition (PSPR) to the Electrocardiogram (ECG) signals. PSPR is used for online feature extraction while they apply random forest (RF) to classify ECG data. The proposed pipeline is based on Spark's ML library (MLlib)} \cite{meng2016mllib}, \hl{hence allows other anomaly detection techniques included within MLlib.}

\citet{nitsche2019development} \hl{propose a data science pipeline that is optimized for text classification. The authors benchmark different GPUs to evaluate the performance of their hardware setup which consists of 10 NVIDIA Quadro P6000 and the effect of the number of GPUs on the image processing time. They apply the Naive Bayes classifier that is included in scikit-learn API and achieve above 90\% accuracy on Deutsche Presse-Agentur (dpa) dataset.}

\citet{shaikh2017end} \hl{focus the challenges of ensuring policy fairness within end-to-end ML pipelines. They claim the ML-based tasks are done by engineers that have a variety of professions including data creators and future engineers. Hence, each step of the ML pipeline might be subjected to a policy violation. The authors provide an end-to-end ML pipeline that is based on log management to prevent these violations as manually ensuring policy fairness is highly resource-consuming.}

\citet{boovaraghavan2021mliot} \hl{propose an adaptive end-to-end ML system for IoT applications. Their pipeline is optimized for activity recognition-based tasks including object recognition. Authors claim that the main challenge regarding end-to-end pipeline is due to the heterogeneity of IoT applications. Authors evaluate their pipeline with various hardware platforms and datasets while comparing prediction time and accuracy per each machine learning technique they applied.}

\citet{molinara2020end} \hl{propose an end-to-end ML-based indoor air monitoring system for contaminant classification. Authors compare the performances of Multi Layer Perceptron (MLP) to CNN and LSTM based deep learning techniques while testing the performance of MLP and CNN on ESP32 MCU. They investigate the power consumption of the MCU regarding the utilized ML technique. They claim the proposed system is only lacked to classify alcohol and acetone due to their chemical similarities.}

\citet{vinzamuri2020end} \hl{propose an end-to-end context-aware anomaly detection system that requires time-series data. The proposed system utilizes a semi-supervised algorithm with Sparse Gaussian Graphical Models. They benchmark the pipeline on several public datasets. The authors claim semantics can improve the Gaussian Graphical Models further beyond other anomaly detection techniques. Their ML comparison is based on F-Score as the authors mention that the proposed pipeline is promising for industrial IoT environments.}

\citet{li2020pyodds} \hl{develop an end-to-end automated anomaly detection system. They utilize Apache Spark backend server to run the query-based operations. After the user provides a dataset, the proposed system automatically selects the most appropriate algorithm then applies anomaly detection. Finally, the results are shown in figures within the pipeline. They benchmark the proposed system based on several datasets while applying quantification analysis.}

\hl{Our pipeline utilizes scikit-learn and TensorFlow for the anomaly detection while relying on TensorFlow Lite for the deployment on the fog, TensorFlow Lite Micro for the deployment on the edge devices. Hence, it allows the lightweight implementation of anomaly detection techniques while offering a variety of communication protocols (e.g., Bluetooth low energy (BLE)) and sensor types for the inference. The Table} \ref{tab:relatedWorkComparison} \hl{compares our pipeline with the previous works based on significant features that determine the efficiency of the pipeline.}

\begin{table}[!t]
\centering
\footnotesize
\caption{\hl{Comparison of AnoML-IoT with the Previous Works}}
\label{tab:relatedWorkComparison}
\begin{adjustbox}{scale = 0.73, center}
\def\arraystretch{1.2}
\begin{tabular}{@{}lccccccc@{}} 
\toprule[1px]
Related Work                                             & Topic             & Environment & Commercial & Open-source & End-to-End & Time-series Data & Adaptability  \\ 
\midrule[0.5px]
Azure Machine Learning Pipeline \cite{nilspohlmann_create} & General           & General~    & \boldcheckmark        &           & \boldcheckmark        & \boldcheckmark              &             \\
AWS Machine Learning Pipeline \cite{aws_build_2021}   & General           & General     & \boldcheckmark        &           & \boldcheckmark        & \boldcheckmark              & \boldcheckmark           \\
\citet{prado2020bonseyes}               & Classification    & IoT         & \boldcheckmark        &           & \boldcheckmark        & \boldcheckmark              &             \\
\citet{sutton2018physonline}            & Anomaly Detection & Medical     &          & \boldcheckmark         &          & \boldcheckmark              &             \\
\citet{nitsche2019development}          & Classification    & Linguistics &          &           &          &                &             \\
\citet{shaikh2017end}                   & Policy            & General     &          &           & \boldcheckmark        & \boldcheckmark              &             \\
\citet{boovaraghavan2021mliot}          & Classification    & IoT         &          &           & \boldcheckmark        & \boldcheckmark              &             \\
\citet{molinara2020end}                 & Classification    & IoT         &          &           & \boldcheckmark        &                &             \\
\citet{vinzamuri2020end}                & Anomaly Detection & IoT         &          &           & \boldcheckmark        & \boldcheckmark              &             \\
\citet{li2020pyodds}                    & Anomaly Detection & General     &          &           & \boldcheckmark        & \boldcheckmark              & \boldcheckmark           \\
AnoML-IoT                                                & Anomaly Detection & IoT         &          & \boldcheckmark         & \boldcheckmark        & \boldcheckmark              & \boldcheckmark           \\
\bottomrule[1px]
\end{tabular}
\end{adjustbox}
\end{table}

\section{The Architecture of the AnoML-IoT Pipeline}
\label{section:Architecture}

In this part, we present the overall architecture of our pipeline by defining main components, describing the workflows that differ according to application scenario, and explaining how to automate these workflows to maintain the pipeline. 

\subsection{AnoML-IoT Layers and Application Scenarios}

IoT infrastructures should provide semantic data exchange to be considered as completely ubiquitous. Current technologies that are utilized in IoT architectures are rapidly evolving to achieve semantic interoperability, hence causing the debate of what kind of infrastructure is needed for a certain task. Due to each technology has a variety of pros and cons per IoT environment, extensive testing is required to decide on IoT elements (e.g., wireless technologies, edge/fog node types, sensors, and actuators). Building an IoT application from scratch to perform these tests requires intensive labor. Thus, reconfigurable IoT sensing architecture that allows the implementation of various ML-based application scenarios including anomaly detection with minimum user (e.g., data scientist) interaction is required. AnoML-IoT allows the implementation of a variety of application scenarios where the scenarios are evolved around the platform that the anomaly detection is implemented. In this context, AnoML-IoT consists of three layers: 

\begin{itemize}

  \item \textit{Edge Layer}: The edge layer contains edge nodes that consist of microcontrollers, sensors, and actuators. These nodes are physically observing the IoT environment by gathering data via sensors while conducting physical operations via actuators (e.g., the fan stops working when a certain degree is reached).
  
  \item \textit{Fog Layer}: The fog layer contains fog nodes that get the data from edge nodes and process it according to the end user's preferences. Usually, fog nodes offer more computing power than edge nodes while allowing flexible deployment options. Even though fog computing may be considered as an alternative to cloud computing, fog nodes can also act as an IoT gateway between the edge nodes and the cloud. In this work, we utilize Raspberry Pi 4 as a fog device and present our results. We believe, similar Linux-based devices can be used as fog devices instead of the Raspberry Pi for the proposed pipeline. 
  
  \item \textit{Cloud Layer}: The cloud layer provides services ranging from data management, storing, applying anomaly detection to developing multi-purpose frameworks. Even though, cloud computing offers many benefits (e.g., automatic service integration, and high accessibility), edge/fog computing is preferred where time-critical or confidential applications (e.g., industrial) are present to reduce reliance on cloud services. While the edge and fog layer might contain standalone nodes, the cloud layer requires interaction with other layers.
  
\end{itemize}

\begin{figure}[!t]
  \centering
  \includegraphics[scale = 0.67]{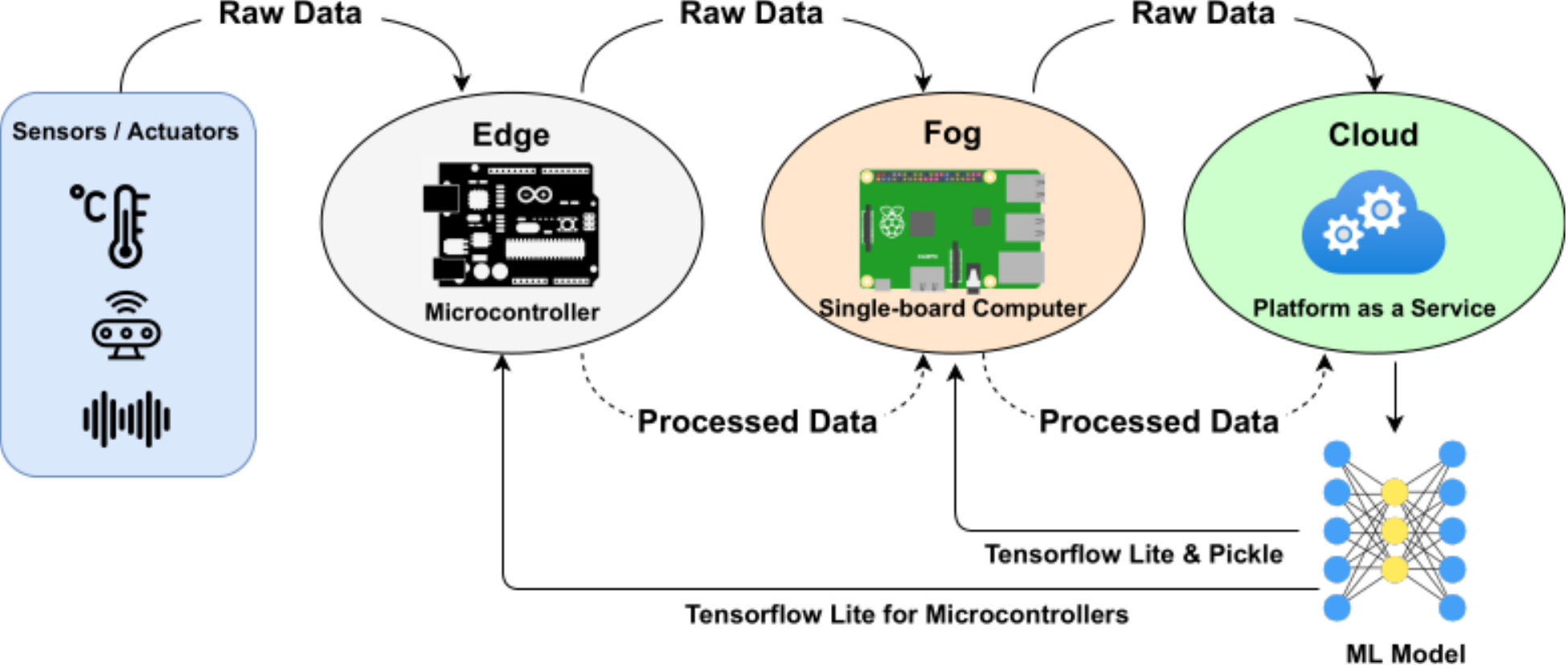}
  \caption{\hl{Overview of the data circulation and the pipeline application scenarios based on the location of anomaly detection: (i) If the anomaly detection is done on the cloud, the edge and fog devices only forward data, (ii) if the anomaly detection is done on the fog, the edge sends raw data to fog, then the fog might send processed data to the cloud, (iii) if the anomaly detection is done the edge, the processed data might be sent to fog and then to cloud. Each scenario has its pros and cons that we introduce in section} \ref{Sec:evaluation2}. \hl{Here processed data contains information to be used to decide if the data is anomalous or not. Thus, it might be either binary (e.g., 0 for normal data, 1 for anomalous data) or a decimal(float) as an anomaly score that varies in the range of -1 to 1. The model in the cloud will be updated with the new normal data according to the predetermined intervals.}}
  \label{fig:applicationScenarios}
\end{figure}

Figure \ref{fig:applicationScenarios} summarizes the working principle of the machine learning pipeline. The anomaly detection can be done within the pipeline on edge, fog, or cloud. The edge and fog devices also can be utilized just to forward raw data to the platform at one upper level. The initial training requires prior data. Hence, if there is no dataset to be used for initial training, the edge and fog will send only raw data until the cloud can generate an efficient ML model. Then the model will be deployed to edge, fog, or cloud. According to the given intervals, the new model will replace the old model to keep the system up-to-date. Having an up-to-date ML model is significant to prevent a decrease in efficiency that depends on the dynamic context.

\subsection{The AnoML-IoT Open-Source Tools}

AnoML-IoT consists of three main tools: (i) Edge to Cloud Code Generator (ECCG), (ii) Node-RED package, (iii) python library. Now we introduce what these tools provide, what are their functionalities, use cases, and roles within the AnoML-IoT pipeline. The source code of the tools is published in GitLab\footnote{https://gitlab.com/IOTGarage/anoml-iot-analytics}.

\subsubsection{The Edge to Cloud Code Generator} 

The ECCG is a web-based code generator that generates edge code to be either used for inference or to transmit data to the fog node. Currently, it is only compatible with Arduino IDE \cite{fezari2018integrated}. It has a user-friendly interface, where even the data scientist with minimal IoT knowledge can design a basic IoT application that contains several sensors where the sensor data can be transferred between layers. The characteristics of ECCG including the justification for design choices are given below:

\begin{itemize}

\item Currently, five sensor types are available: \textit{temperature}, \textit{humidity}, \textit{air quality}, \textit{light}, \textit{loudness}. The end-user can simultaneously select all sensors. Selected sensor data will be included within the pipeline. \textit{Justification}. These six sensor types are among the most common sensors that are utilized in IoT applications. The generated code clearly describes how the sensor data is received, and processed, thus allowing the easy adaptation of the code for a similar type of sensor.

\item Four communication protocols are available: \textit{Wi-Fi} \cite{hiertz2010ieee}, \textit{Bluetooth Classic} \cite{chang2014bluetooth}, \textit{BLE} \cite{heydon2012bluetooth}, and \textit{Zigbee} \cite{ergen2004zigbee}. The end-user can only select one. The application also includes additional settings regarding communication protocols for advanced users. \textit{Justification}. The selected four communication protocols occupy the vast majority of the IoT market and offer a variety of topologies (e.g., mesh, star, tree). Understanding the basics of how to establish these wireless technologies enables the implementation of high-range IoT applications. The generated code by ECCG defines how to handle the required network elements (e.g., MAC address, personal area network (PAN) ID). An advanced user may conveniently adapt the generated code to be used with other IoT communication protocols that are out-of-scope of this project such as WirelessHART \cite{song2008wirelesshart}.

\item The data transfer rate determines the time between two data blocks. It is in milliseconds that ranges from 30000 to 300000. After copying to code the user can set the data transfer rate as desired. However, we do not recommend setting it below 30 seconds as it is the time that is required for the module to initialize. \textit{Justification}. Data generating time differs per sensor module. Controlling the data transfer rate is necessary to ensure the integrity of transmitted data.

\item The sensor locations are determined by the end-user. Identification (ID) number starting from 00 is given per location. The application supports up to 99 locations. \textit{Justification}. Data scientists work with comma-separated value (CSV) files, to handle the further progressing of the data. The data in CSV files mostly in pairs as \textit{text-value} where \textit{text} is the identifier, and \textit{value} is digital presentation of a physical quantity. To generate such files, data is transmitted in data-serialization formats (e.g., JavaScript Object Notation (JSON), Extensible Markup Language (XML)) supported by IoT application standards. Thus, the ECCG allows an end-user to define location. Then, it generates a unique location ID number to be used in data transmission.

\item Three microcontroller types are available: Arduino Nano 33 BLE Sense, Arduino Nano RP2040 Connect, and Raspberry Pi Pico. The end-user can select one of the microcontrollers to obtain the edge node ID number to be used to identify the microcontroller types when needed. \textit{Justification.} The ECCG supports the top three microcontrollers that are officially supported by TensorFlow Lite for Microcontrollers (TFLM) \cite{warden2019tinyml}. Supporting a variety of microcontrollers allows users to design a heterogeneous IoT application, where the environment benefits from different features of these devices (e.g., RAM, flash memory). 

\item The ECCG allows sending lowest, mean, or highest sensor data which are generated during the time interval determined by the user. \textit{Justification}. In some cases, the normal range might just be determined by lower or upper limits. Hence, we allow the user to decide on the data to be sent. If the user selects mean, the mean of the number of data points generated by the sensor during the predetermined time interval will be sent.

\end{itemize}

\begin{figure}[!t]
  \centering
  \includegraphics[scale = 0.40]{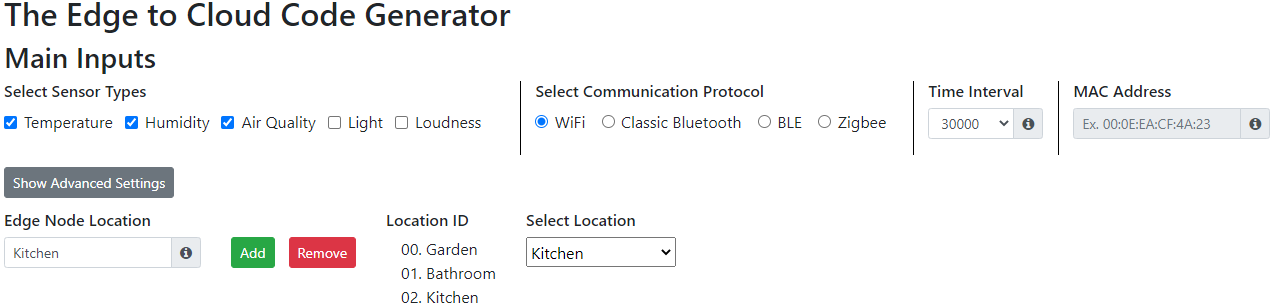}
  \caption{\hl{Illustrates the main input section of the ECCG. Each input is strictly controlled for the following reasons: (i) to prevent cross-site scripting attacks, (ii) to prevent bugs that may occur in the code due to mistyping. Each text input has a tooltip that clarifies what kind of information should be given. In addition, placeholders demonstrate an example input.}}
  \label{fig:mainInputs}
\end{figure}

The user interface of the ECCG facilitates usability by navigating users via input controls. Figure \ref{fig:mainInputs} demonstrates the main input section of the code generator while the example inputs are given. Minimalist design is preferred to assist a data scientist with no prior IoT knowledge. Thus inputs are controlled, tooltips are included, and example inputs are given as placeholders.

\begin{figure}[!h]
  \centering
  \includegraphics[scale = 0.55]{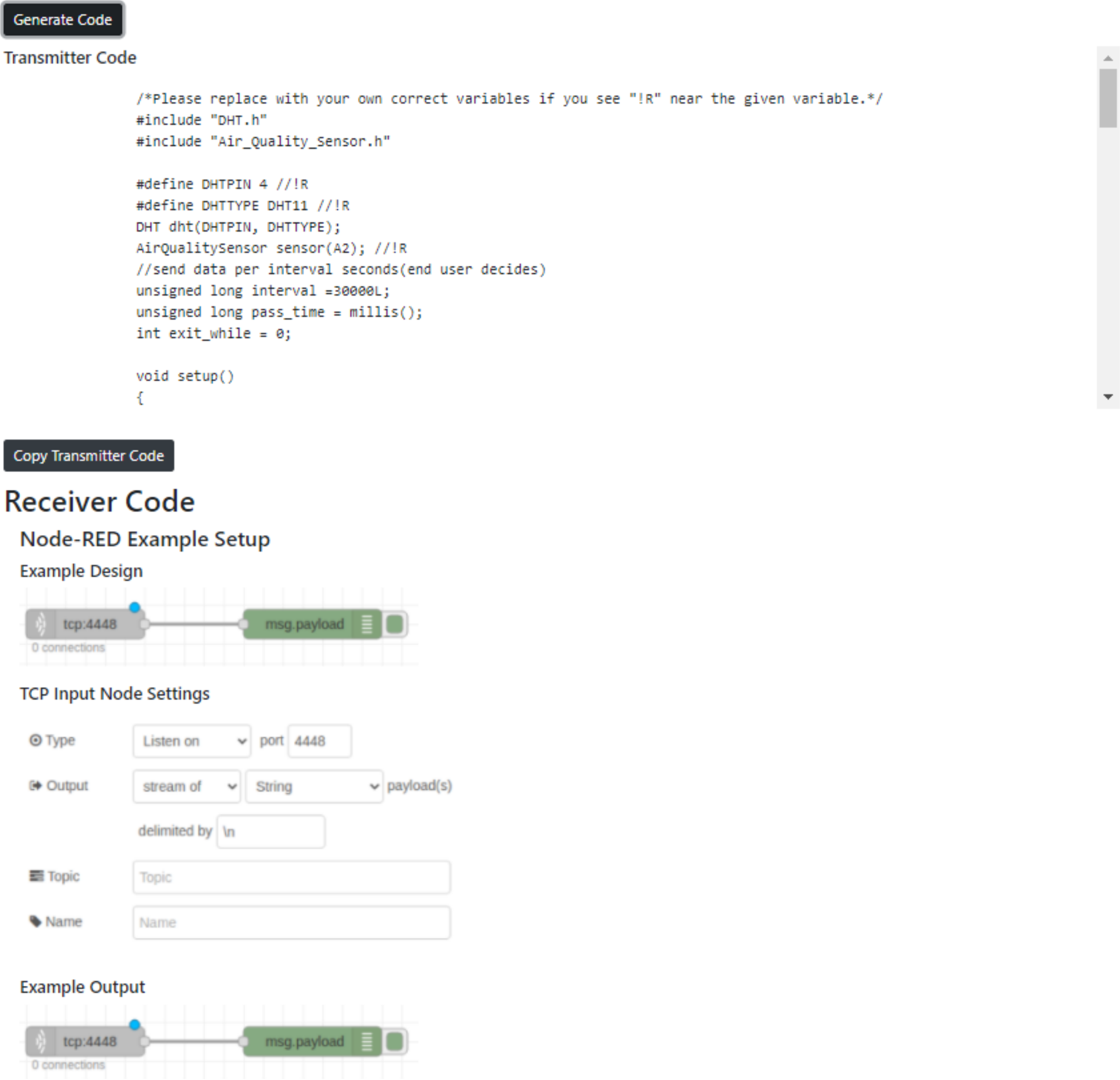}
  \caption{\hl{Illustrates the generated code block and the example Node-RED setup. In the example scenario, Raspberry Pi Pico is selected as an edge node/transmitter. The generated transmitter code is written in C++ and should be deployed via Arduino IDE. The Raspberry Pi 4B is acting as a fog node/receiver.}}
  \label{fig:ECCGGeneratedCode}
\end{figure}

After the inputs are given, the final step is clicking on the "Generate Code" button. The ECCG will output the followings: (i) the edge code that is ready to be deployed to the microcontroller via Arduino IDE, (ii) the Python3 script that should be run on the fog device, (iii) the Linux commands to be run via terminal, (iv) example Node-RED setup. The outputs differ according to the given inputs. For example, the Linux commands are only required if the Bluetooth Classic will be used within the pipeline. The Figure \ref{fig:ECCGGeneratedCode} illustrates how the generated code blocks are presented. Two main buttons are included for each code block: the first button generates the code shown in the pre-scrollable division to let a data scientist examine the codes before further progression, the second button copies the code without breaking the format to prevent possible errors. The user manually uploads the code into the microcontroller.

\hl{We assume the following scenario: the edge node (micro-controller) gathers physically observed data via sensors and sends it to a fog node(a single-board computer (SBC)) where the data is either processed or forwarded to the cloud via Node-RED. Thus, the ECCG consists of three main sections: (i) the input field where the end-user determines the basic characteristics of the desired IoT application, (ii) the transmitter field where the edge node code is generated for a microcontroller, and (iii) the receiver field where the fog node code is generated for a single-board computer. Table} \ref{tab:ECCGspecs} \hl{illustrates the specifications of the ECCG in detail.}

\begin{table}[!t]
\centering
\small
\caption{The Specifications of The Edge to Cloud Code Generator}
\label{tab:ECCGspecs}
\begin{adjustbox}{scale = 0.75, center}
\begin{threeparttable}
\renewcommand{\arraystretch}{1.1}
\begin{tabular}{@{}lll@{}}
\toprule[1pt]
\multirow{11}{*}{Main Inputs}      & \multirow{5}{*}{Sensor Types}            & Temperature                   \\
                                   &                                          & Humidity                      \\
                                   &                                          & Loudness                      \\
                                   &                                          & Light                         \\
                                   &                                          & Air Quality                   \\ \cline{2-3} 
                                   & \multirow{3}{*}{Communication Protocols} & Wi-Fi                          \\
                                   &                                          & Bluetooth Classic             \\
                                   &                                          & BLE                           \\
                                   &                                          & Zigbee                        \\ \cline{2-3} 
                                   & \multirow{3}{*}{Edge Node Types}         & Arduino Nano 33 BLE Sense     \\
                                   &                                          & Arduino Nano RP2040 Connect      \\
                                   &                                          & Raspberry Pi Pico             \\ \cline{2-3} 
                                   & \multirow{2}{*}{Edge Node Identifiers}   & Edge Node Location            \\
                                   &                                          & Edge Node ID Number           \\ \hline
\multirow{10}{*}{Advanced Inputs}  & \multirow{4}{*}{Wi-Fi}                    & Service Set Identifier (SSID) \\
                                   &                                          & Password                      \\
                                   &                                          & Host IP Address               \\
                                   &                                          & Host Port                     \\ \cline{2-3} 
                                   & \multirow{3}{*}{Bluetooth \& BLE}               & *MAC Address                   \\
                                   &                                          & Module Name                   \\
                                   &                                          & Module PIN                    \\ \cline{2-3} 
                                   & \multirow{3}{*}{Zigbee}                  & PAN ID                        \\
                                   &                                          & Destination Address High      \\
                                   &                                          & Destination Address Low       \\ \hline
\multirow{7}{*}{Generated Outputs} & \multirow{3}{*}{Node-RED Example Setup}  & Node Settings                 \\
                                   &                                          & Workspace Design              \\
                                   &                                          & Example Functions             \\ \cline{2-3} 
                                   & \multirow{2}{*}{Transmitter Code}        & Arduino Scripts               \\
                                   &                                          & Python Scripts                \\ \cline{2-3} 
                                   & \multirow{2}{*}{Receiver Code}           & JavaScript Function Nodes     \\
                                   &                                          & Python Scripts                \\ \hline
\end{tabular}
\begin{tablenotes}
\small
\item[*] Among advanced inputs, only the MAC address is obligatory. The default options that are included in the code are explained via comments in detail.
\end{tablenotes}
\end{threeparttable}
\end{adjustbox}
\end{table}

\subsubsection{The Node-RED Package}

Node-RED \cite{Node-RED} is an open-source browser-based workflow development tool that runs on Node.js \cite{tilkov2010node} based runtime. We can develop workflows via utilizing drag-and-drop nodes on resource-constraint environments such as Raspberry Pi thanks to the non-blocking nature of Node.js. Hence, we use Node-RED on the fog device to develop and control the workflows within the AnoML-IoT pipeline. Even though Node-RED contains many open-source packages developed for the IoT networks, we could not find any up-to-date package that provides the development of Bluetooth Classic, BLE, Zigbee, and Wi-Fi networks while offering customization choices. Hence, we developed and published our package under the name of "node-red-contrib-IoT-procotols". The package contains the following nodes:

\begin{itemize}
    
    \item \textbf{BLE Scanner}. \hl{BLE devices can operate in four different roles: broadcaster, observer, central, and peripheral. The task of the BLE Scanner node is to scan for the advertising BLE peripherals. It can specifically search for a local name and output one of the followings: whole peripheral as an object, MAC address, and advertisement data. So, the user can use this node either to discover the MAC address of the target BLE peripheral or to get the advertisement data.}  
    
    \item \textbf{BLE Connect}. \hl{This node establishes a BLE connection with the target peripheral device. While one peripheral device can only be connected to one central, central devices can manage multiple peripherals. Hence, our pipeline allows the deployment of multiple edge devices via BLE while the maximum number of peripherals that can be connected depends on the system-on-a-chip (SoC) of the microcontroller. For example, the Arduino Nano 33 BLE Sense} \cite{arduinoBLE} \hl{SoC nRF52840} \cite{nrf52840} \hl{supports up to 20 parallel connections.} 
    
%
    
\end{itemize}



\subsubsection{The AnoML.py and SetupAnoML.sh}

\textbf{AnoML.py}. \hl{Library of functions in python which can support most of the tasks in developing ML models with various types of normalization and data points. Generating different models for the cloud, fog and edge platforms using various normalization methods and data points is a labor-intensive task. Evaluating different models for unsupervised ML when there is a small amount of anomalous data is also another challenge. Our Python library supports data preprocessing to generating ready to deploy ML models for the cloud, fog, and edge for unsupervised ML. It also provides performance visualization for each data point while including normalization techniques which helps to select the most efficient model. The proposed library provides data cleaning, normalization, reduction, scaler, and visualization functions before feeding into machine learning models. The library also provides various functions that allow training and testing. The inference part of the library provides functions to evaluate different machine learning models generated using our library at both fog and cloud platforms so the performance of the platforms can be compared.}

\textbf{SetupAnoML.sh}. \hl{It acts as an installer to prepare both fog and cloud platforms for action. It installs correct versions of necessary packages and libraries required to run inference. Once the prerequisites are handled, it can then download, install and configure packages such as Node-RED and TensorFlow runtime. Also, it sets our custom-developed services to listen on web ports to order to allow microcontrollers to interact with fog devices and fog devices to interact with the cloud platform. The script can be configured to download files and configurations either from Google Drive or a local/remote FTP Server. Users also can manually download, install and configure all prerequisite requirements and model files if it is desired.}

\section{The Data Circulation}
\label{section:dataCirculation}

\hl{The nature of the input data shapes the characteristics of the data science pipeline. Hence, gathering raw data is the first step in this kind of pipeline. In the IoT environment, the sensors generate a variety of data in various formats. While most of the temperature sensors generate data in floating-point numbers (floats) to provide more accuracy, the light sensors that measure light density usually provide data in integer. This is significant for two main reasons: (i) each data type occupies memory in different sizes, (ii) there might be different regulations} \cite{8766229, yao2005applying} \hl{according to the application domain for certain data types such as floats. Hence, the memory usage should be optimized for microcontrollers that have very limited memory. The memory usage might also differ according to the microcontroller architecture. For instance, while Arduino Uno} \cite{badamasi2014working} \hl{stores int values in two bytes, Arduino Due} \cite{due2017arduino} \hl{stores int values in four bytes. The data circulation within the AnoML-IoT pipeline consist of four main steps: (i) data ingestion, (ii) data training, (iii) model deployment, (iv) inference and maintaining.}  

\subsection{Data Ingestion}

\hl{Data ingestion is the first step of all kinds of data science pipelines unless the user already has a ready-to-deploy ML model. In our pipeline, the ECCG is the main tool that utilizes the data ingestion process. Figure} \ref{fig:mainInputs} \hl{illustrates the choices offered by the ECCG. The user can select among the offered choices to obtain a generated code. The generated code can be copied without disrupting the format. Multiple communication protocols and Node-RED example that demonstrates how to receive data via fog device are also provided.}

\begin{figure}[!b]
  \centering
  \includegraphics[scale = 0.49]{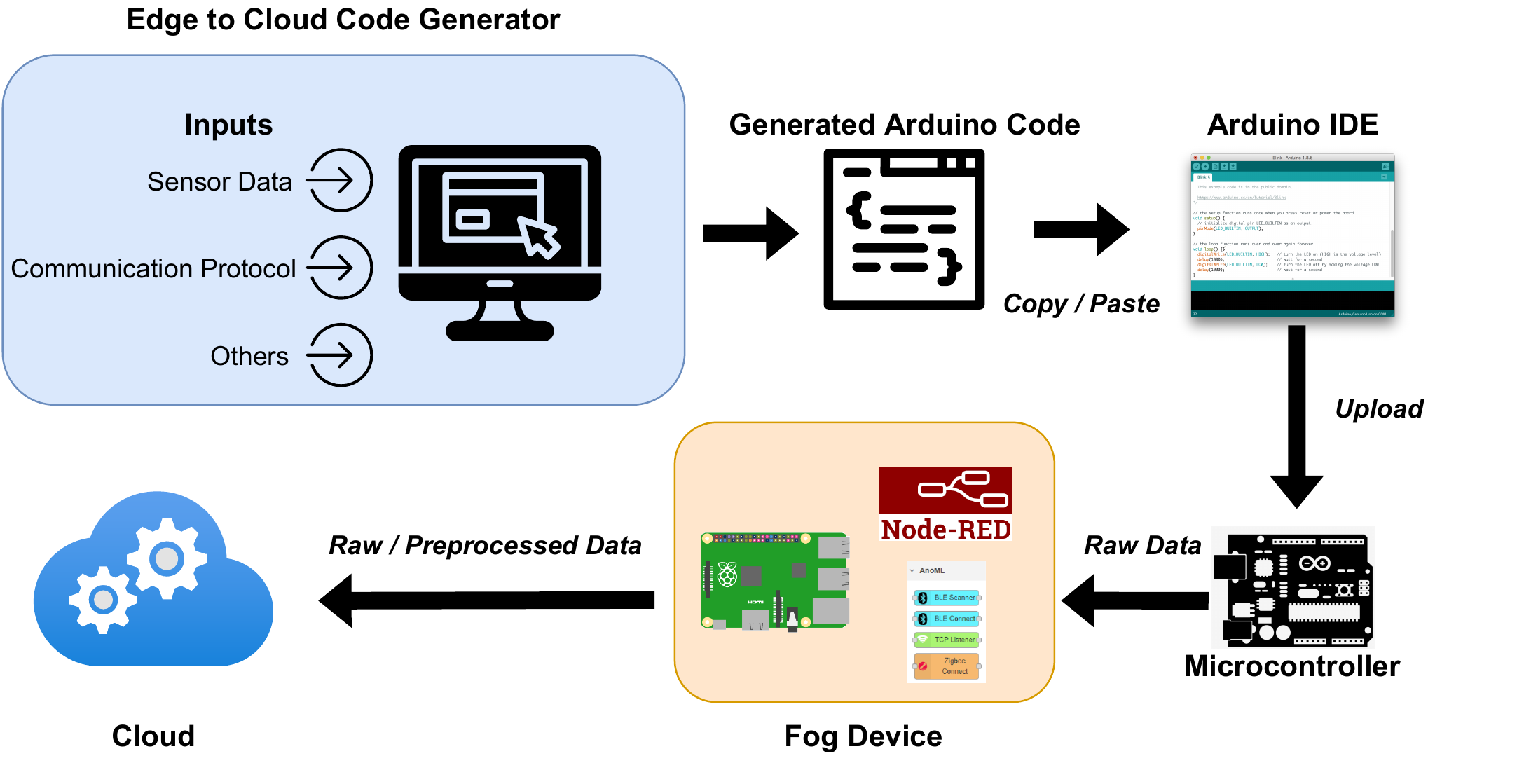}
  \caption{\hl{Illustrates data ingestion phase. The data preprocessing can be done on all platforms. However, due to the limited computing power of edge and fog devices, we prefer to utilize the cloud. In addition, if more than one edge device is connected to the fog device, data preprocessing can be done on the fog device to identify the edge devices.}}
  \label{fig:dataGathering}
\end{figure}

\hl{What the ECCG generates is an example edge code. Besides, in our setup, we use Grove sensors and shields} \cite{grove} \hl{while utilizing Raspberry Pi 4B as a fog device. The IoT environment is very heterogeneous hence, the user might have or want to use different sensors, microcontrollers, or fog devices. The generated code by the ECCG should require minimal editing even when this is the case as we also provide instructions via comments for the exact lines within the code that might require editing due to individual preferences. If needed, the user can edit the generated code after copying it from the ECCG to Arduino IDE. Then, the code can be uploaded to a microcontroller. To complete these tasks the user needs a computer that can access the internet and run Arduino IDE. So, the tasks of copying the code and uploading it to a microcontroller are handled manually. Figure} \ref{fig:dataGathering} \hl{demonstrates the data ingestion process.}

\subsection{Model Training}

\begin{figure}[!t]
  \centering
  \includegraphics[scale = 1.00]{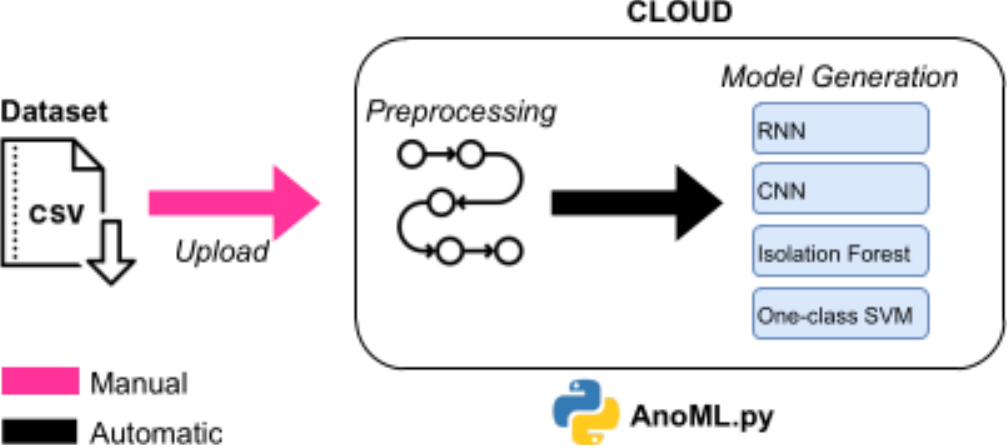}
  \caption{\hl{Demonstrates the model training phase. Currently, the built-in algorithms within the AnoML.py are RNN, CNN, isolation forest, and One-class SVM. Multiple datasets can be utilized at the same. The storage place of these models depends on the preferences of the user. During our evaluation we used Google Drive} \cite{quick2014google} \hl{to store our models as it can be easily integrated into Google Colab} \cite{carneiro2018performance}. \hl{Another option would be running an FTP server. The user also can utilize both at the same time.}} 
  \label{fig:modelTraining}
\end{figure}

\hl{Model training can be done before ingestion if the user already has a training dataset. Otherwise, the user should follow the steps mentioned in the data ingestion phase. The AnoML.py library contains all the required functions that are needed to generate a ML model. Uploading the dataset to a cloud platform is handled manually. After uploading the dataset, the user should preprocess the data to convert to an appropriate format for the ML algorithms. Then, the ML models are generated based on the user preferences. User can generate multiple ML models at the same time. Here, we facilitate the model training by providing user a python library that is capable of preprocessing the time series data and generating multiple ML models based on given parameters with only a few line of codes. The required time depends on the capabilities of a cloud platform that is used during the training. The user should decide on the followings: (i) the size of the training dataset, (ii) the algorithm specific parameters(e.g., contamination factor for Isolation Forest). Figure} \ref{fig:modelTraining} \hl{illustrates the model training phase.}

\subsection{Model Deployment}

\begin{figure}[!t]
  \centering
  \includegraphics[scale = 0.80]{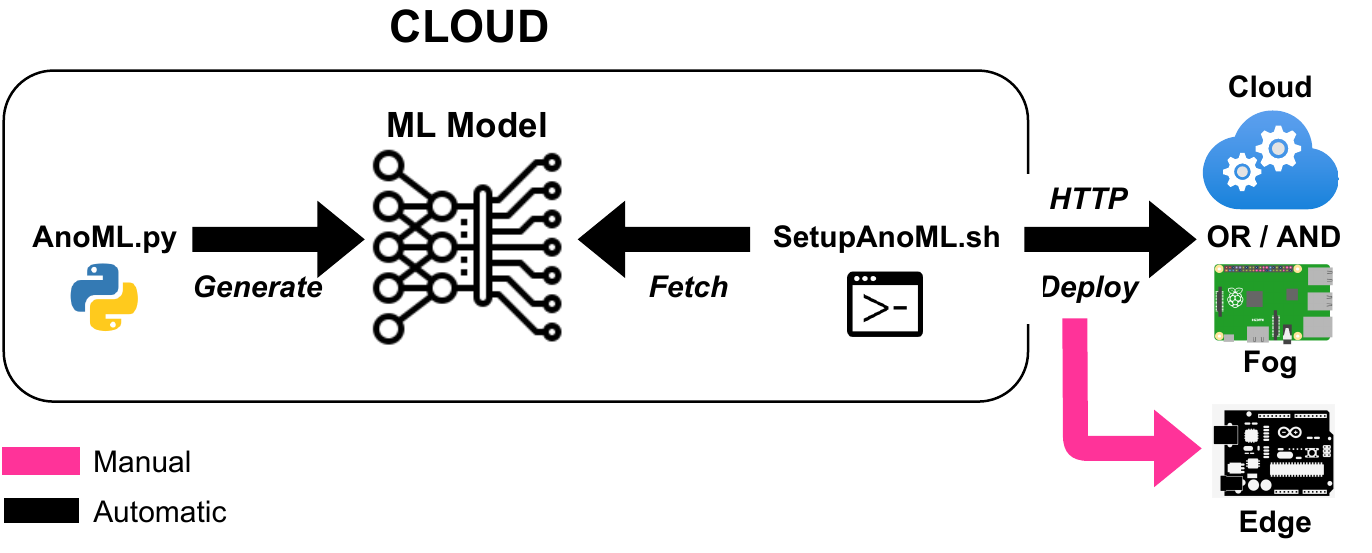}
  \caption{\hl{Illustrates the model deployment phase. SetupAnoML is an executable Linux shell script that does the followings in order: (i) fetches generated ML models from Google Drive or FTP server, (ii) installs required libraries and packages, (iii) deploys models over HTTP. The deployment to edge should be handled manually. We are considering to provide automated edge ML deployment function via over-the-air (OTA) transmission in the future versions of AnoML-IoT.}}
  \label{fig:modelDeployment}
\end{figure}

\hl{The next step after training the model is model deployment. Here we provide an executable shell script SetupAnoML.sh that facilitates the model deployment phase by automating several key tasks. The user should provide the following inputs to the script: (i) the platform (edge or cloud) where the anomaly detection technique will be deployed, (ii) the details of the place where the model is stored (e.g., server, username, password, and port names for an FTP server, Google Drive token or Google Drive). After these inputs are given, the script will automatically download, install, and configure all the required software packages (e.g., libraries, models, configurations). Then the script automatically will deploy the models to cloud, fog, or both cloud and fog platforms. The user can deploy multiple models at the same time to different platforms. Figure} \ref{fig:modelDeployment} \hl{demonstrates the model deployment process.}

\subsection{Inference and Maintaining}
\hl{Successful deployment results in inferring all models on each type of platform. We developed an end-to-end pipeline which allows a user to generate ML  models for anomaly detection from sensor data at all platforms. We developed a performance monitor which can present visualization of performance and accuracy of all type platforms, ML models, data-points and normalization/reduction techniques. Maintenance of data is a key aspect for evolution, we recommend that user should make it a practice to visualize performance comparison as a process of decision making in order to enhance performance and accuracy on all platforms.}

\section{Evaluation}
\label{Sec:evaluation2}

\begin{figure}[!t]
  \centering
  \includegraphics[scale = 0.22]{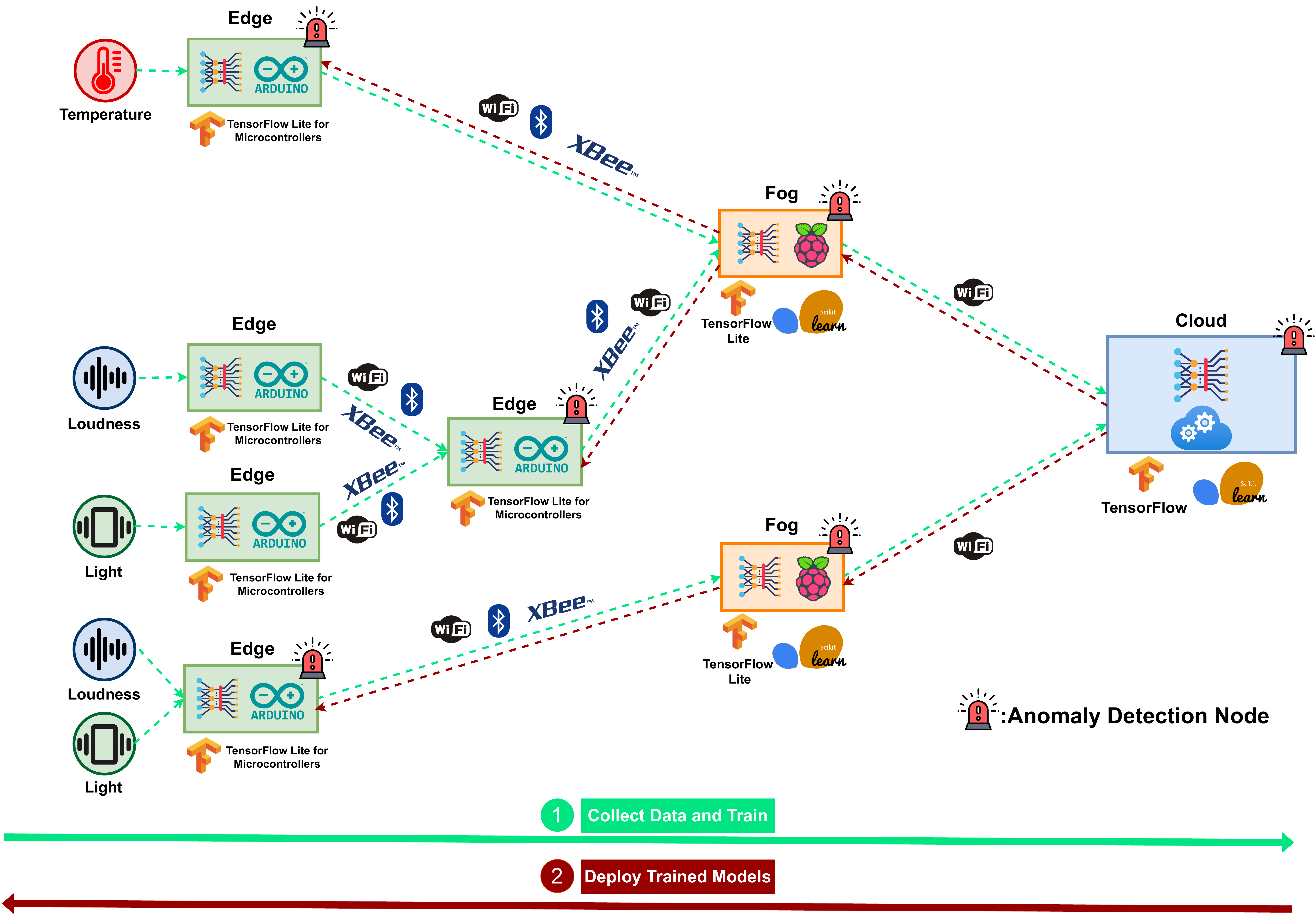}
  \caption{\hl{Illustrates the possible scenarios that can be generated within the AnoML-IoT pipeline. We allow the use of various topologies, wireless communication protocols, and execution of use case scenarios shaped around the decision of anomaly detection platform.}}
  \label{fig:testbedHK}
\end{figure}

In this section, we present the evaluations that are done to measure the efficiency of the pipeline with various configurations enabled. A data science pipeline evaluation is not a trivial task if the environment is heterogeneous such as IoT. Figure \ref{fig:testbedHK} demonstrates possible scenarios available within our pipeline where different sensors and wireless communication protocols are used.

\subsection{Wireless Protocol Comparison}

The decision on the wireless protocol selection depends on the application domain as each protocol has its benefits and disadvantages. There are several important features to be considered when designing an anomaly detection pipeline for the IoT environments:

\begin{itemize}

    \item \textbf{Latency}. The time that takes for one network package to be transmitted from the transmitting endpoint to the receiving endpoint. Latency optimization is very significant to achieve near real-time processing and inference. It is one of the significant features that determine the anomaly detection time. Low latency is aimed for the applications (e.g., industrial) where the anomaly detection time is critical \cite{giraldo2018survey}.
    
    \item \textbf{Power consumption}. IoT is a resource-constraint environment, hence power consumption is one of the significant features to be considered.
    
    \item \textbf{Network Topology}. The data circulation and the system design depend on the network topology. While there are many topologies (e.g., star, mesh, ring) offered by the current communication protocols, the mesh, and star networks are the most common ones that are established in IoT environments.
    
\end{itemize}


Table \ref{tab:protocolComparison} compares the wireless communication protocols utilized in the proposed pipeline based on the aforementioned features. We measured the latency by taking the mean of the passed time for the 1000 transmitted packages. Edge and fog devices were placed next to each other during the tests hence there was no physical barrier between the devices. No packet drop is observed. While we observed similar latency for the BLE, Wi-Fi, and Zigbee, the Bluetooth Classic had the highest latency. However, it is challenging to have a conclusion as many factors that might affect the latency. Hence, to get the most realistic results, the tests should run in a real environment where anomaly detection takes place.

\begin{table}[!t]
\footnotesize
\caption{\hl{Wireless Protocol Comparison}}
\label{tab:protocolComparison}
\begin{adjustbox}{scale = 0.95, center}
\begin{threeparttable}
\def\arraystretch{1.2}
\begin{tabular}{@{}lcccccc@{}}
\cline{2-7}
                  & \multicolumn{4}{c}{Latency (ms)}                                  & \multirow{2}{*}{Power Consumption} & \multirow{2}{*}{Topology*} \\ \cline{2-5}
                  & Edge to Edge & Edge to Fog & Fog to Fog & Fog to Cloud            &                                    &                           \\ \hline
Wi-Fi             & 18.24        & 14.566      & 17.25      & 21.23                   & High                               & Star                      \\
Bluetooth Classic & 195.13       & 171.15      & 187.15     & NA                      & Medium                             & Point-to-Point            \\
BLE               & 11.23        & 13.45       & 13.21      & NA                      & Low                                & Mesh                      \\
Zigbee            & 18.56        & 16.66       & 14.56      & NA                      & Low                                & Mesh                      \\ \hline
\end{tabular}
\begin{tablenotes}
\setlength\labelsep{0pt}
 \small
 \item NA: Not Applicable. $\star$: The most common topology is given.
\end{tablenotes}
\vspace{-0.25cm}
\end{threeparttable}
\end{adjustbox}
\end{table}

\subsection{Dataset and Testbed}

\hl{Realistic datasets and testbeds are required to achieve the most promising results. To evaluate the efficiency of the pipeline, we built a testbed that contains components that are low-cost and accessible to most of the IoT community. When designing the IoT testbed, the sensor selection is the initial task that shapes the main futures of the testbed. In cyber-physical environments, there are several behaviors (e.g., temperature, noise, humidity) that can be considered as common. Hence, we observe these common behaviors via our testbed and generate a dataset. Table} \ref{tab:testbed} \hl{demonstrates the components utilized during the evaluation.}

\begin{table}[!t]
\centering
\footnotesize
\caption{\hl{Testbed Components}}
\label{tab:testbed}
\begin{tabular}{@{}ll@{}} 
\toprule[1px]
Sensors               & \begin{tabular}[c]{@{}l@{}}Grove - Temperature \& Humidity Sensor (High-Accuracy \& Mini) v1.0\\Grove - Light Sensor\\Grove - Loudness Sensor\\Grove - Air Quality Sensor v1.3\\Grove - UART Wifi V2\\Digi XBee 3 Zigbee 3 RF Module\\\end{tabular}  \\ 
\midrule[0.5px]
Microcontrollers      & \begin{tabular}[c]{@{}l@{}}Arduino Nano 33 BLE Sense\\Raspberry Pi Pico\\Arduino Nano RP2040 Connect\end{tabular}                                                                                                                                                                                                             \\ 
\midrule[0.5px]
Shields               & \begin{tabular}[c]{@{}l@{}}Grove - Bee Socket\\Grove Shield for Pi Pico V1.0\\Arduino Tiny Machine Learning Shield\end{tabular}                                                                                                                                                                                               \\ 
\midrule[0.5px]
Single-board Computer & Raspberry Pi 4 Model B                                                                                                                                                                                                                                                                                                        \\ 
\midrule[0.5px]
Software Tools        & \begin{tabular}[c]{@{}l@{}}Putty\\Raspberry Pi Imager\\Node-RED\\Arduino IDE 1.8.15\\Visual Studio Code\\Google Colab\end{tabular}                                                                                                                                                                                            \\
\bottomrule[1px]
\end{tabular}
\end{table}

\begin{figure}[!t]
  \centering
  \includegraphics[width = \textwidth]{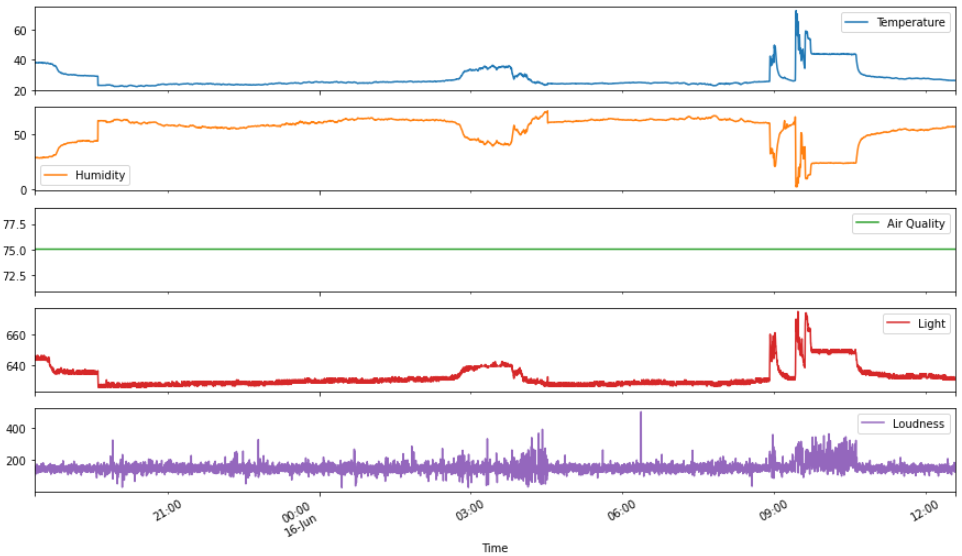}
  \caption{\hl{Demonstrates the behaviour per data type. The controlled anomalies for temperature and humidity data are identifiable. The anomalies in temperature and humidity are created via hair dryer. The air quality, light, and loudness can be considered as faulty due to not responding to our anomaly creation attempts. There might be two reasons for the occurrence of faulty data: (i) the sensors are cheap quality, (ii) these sensors are designed for the AVR }\cite{barrett2012atmel} \hl{architecture. However, during the evaluation we used ARM-based microcontrollers} \cite{bai2015practical}.}
  \label{fig:dataset}
\end{figure}

\hl{We recorded the temperature, humidity, light, loudness, air quality data for around two days. We observed the data at the beginning to form an initial opinion about physical behavioral changes in the test environment. We realized the only temperature and humidity sensors are working as expected and let us creating anomalous behaviors. Hence, we only utilize these two data during the evaluation. Figure} \ref{fig:dataset} \hl{provides visualization of the generated dataset. Arduino Nano RP2040 Connect, Grove sensors, and Putty is used to generate the dataset which is published on Kaggle} \cite{dataset} \hl{where more details are also provided regarding the testing environment. In addition, we utilize the WADI }\cite{ahmed2017wadi} \hl{dataset during the evaluation to analyze how the key ML parameters such as accuracy, F1-score, and prediction time differ according to the ML platform. The dataset contains data from 123 sensors and actuators from a water distribution testbed that had non-stop run for 16 days while being attacked during the last two days. Table} \ref{tab:datasetComparison} \hl{demonstrates the differences between two utilized datasets. By using these datasets, we also evaluate the relationship between the power consumption and data volume.}

\begin{table}[!t]
\centering
\footnotesize
\caption{\hl{Comparison of Datasets}}
\label{tab:datasetComparison}
\def\arraystretch{1.2}
\begin{tabular}{@{}lcc@{}} 
\toprule[1px]
Features        & WADI \cite{ahmed2017wadi}        & AnoML \cite{dataset}   \\ 
\midrule[0.5px]
Amount of Data               & 122,543,744 & 45,906  \\
Number of Rows               & 957,374     & 6,559   \\
Number of Columns            & 129         & 7       \\
Number of Sensors \& Actuators & 123         & 5       \\
Time-Series                  & \boldcheckmark         & \boldcheckmark     \\
Labelled                     & \boldcheckmark         & \boldcheckmark     \\
File Size                    & 588,906 KB  & 265 KB  \\
\bottomrule[1px]
\end{tabular}
\end{table}

\subsection{Anomaly Detection Methods}

We tested the anomaly detection algorithms mentioned in Section \ref{Sec:RelatedWork} on WADI and AnoML datasets, as well as edge, fog, and cloud platforms by applying several scaling/reduction techniques. During the evaluations, we only used the baseline version of anomaly detection algorithms, hence we did not tune the algorithms to get better results to demonstrate the performance of baseline versions. We evaluate the following parameters as they are the core elements that determine the efficiency of an anomaly detection algorithm. \textit{Accuracy} \cite{hasan2019attack} determines the overall correct prediction ratio. \textit{Precision} \cite{davis2006relationship} defines how close the predictions are. \textit{Recall} \cite{davis2006relationship} demonstrates how the anomaly detection algorithm is successful at detecting normal data. \textit{F1-Score} \cite{hasan2019attack} is an evaluation metric that takes class distribution into considering. It might be preferred as the main metric when false detections matter (e.g., industrial environments). \textit{True positive (TP).} The normal data is detected as normal. \textit{True negative (TN).} The anomalous data is detected as anomalous. \textit{False positive (FP).} The normal data is detected as anomalous. \textit{False negative (FN).} The anomalous data is detected as normal. The following equations demonstrate how these parameters are calculated:

    \begin{align*}
    Accuracy &= \frac{TP+TN}{TP+TN+FP+FN} & Precision &= \frac{TP}{TP+FP}\\
    F1-Score &= \frac{2*TP}{2*TP+FP+FN} & Recall &= \frac{TP}{TP+FN} 
    \end{align*}

The TensorFlow has three different APIs. The main API (TensorFlow) contains all the available methods and utilizes high-level API Keras \cite{gulli2017deep} to build neural networks. TensorFlow Lite \cite{tensorflowLite} which is the lightweight version of TensorFlow that is specifically designed for mobile/IoT devices is generated via the main TensorFlow API. TensorFlow Lite for Microcontrollers \cite{GoogleTFMicro2021} only allows a subset of TensorFlow operations to be run on 32-bit microcontrollers. Hence, currently only CNN is supported. TensorFlow Lite models are converted into micro models via TensorFlow Lite converter Python API \cite{tensorflow_converter}. Then, inference is possible at the edge.

Table \ref{tab:wadiEval} and Table \ref{tab:anomEval} \hl{demonstrates the results of anomaly detection tests. We see that baseline CNN at the edge achieves 77\% accuracy and 86\% F1-score. We did not observe any significant differences regarding accuracy and F1-Score. We also see those reduction methods help to reduce the inference time. If we compare scikit-learn methods and TensorFlow methods, we see that scikit-learn performs better when there is more computing power, but TensorFlow might generate better results on fog rather than the cloud. If we use CNN-AE, we see that the accuracy is higher when the standard deviation is applied, but for one-class SVM, we see that skew performs better than the standard deviation. Also, we see that the TF Lite models are significantly less in size than TF models. During the evaluation of the Isolation Forest, the whole dataset must be fit at once, due to nature algorithm. The Raspberry Pi 4B with 4 GB RAM fails due to high RAM usage that occurs because of fitting the whole dataset at once. Hence, we utilized the 64-bit 8 GB version of Raspberry Pi to evaluate the performance of Isolation Forest.}

\begin{sidewaystable*}
\centering
\footnotesize
\setlength{\extrarowheight}{0pt}
\addtolength{\extrarowheight}{\aboverulesep}
\addtolength{\extrarowheight}{\belowrulesep}
\setlength{\aboverulesep}{0pt}
\setlength{\belowrulesep}{0pt}
\caption{Evaluation Results of WADI Dataset}
\label{tab:wadiEval}
\begin{adjustbox}{scale = 0.70}
\begin{threeparttable}
\begin{tabular}{lllcccccccccccccccc}
\toprule[1px]
\multicolumn{3}{c|}{Model Details}                                                                                             & \multicolumn{2}{c|}{Inference Time (ms)}                                                    & \multicolumn{2}{c|}{AUC}                                                                    & \multicolumn{2}{c|}{Accuracy}                                                               & \multicolumn{2}{c|}{Recall}                                                                 & \multicolumn{2}{c|}{Precision}                                                              & \multicolumn{2}{c|}{F1-Score}                                                               & \multicolumn{2}{c|}{Scaling/Reduction Time (s)}                                             & \multicolumn{2}{c}{Model Size (KB)}                                                          \\ 
\hline
Algorithm               & SR                                           & API                                                   & {\cellcolor[rgb]{0.776,0.878,0.706}}Fog      & {\cellcolor[rgb]{0.706,0.776,0.906}}Cloud    & {\cellcolor[rgb]{0.776,0.878,0.706}}Fog      & {\cellcolor[rgb]{0.706,0.776,0.906}}Cloud    & {\cellcolor[rgb]{0.776,0.878,0.706}}Fog      & {\cellcolor[rgb]{0.706,0.776,0.906}}Cloud    & {\cellcolor[rgb]{0.776,0.878,0.706}}Fog      & {\cellcolor[rgb]{0.706,0.776,0.906}}Cloud    & {\cellcolor[rgb]{0.776,0.878,0.706}}Fog      & {\cellcolor[rgb]{0.706,0.776,0.906}}Cloud    & {\cellcolor[rgb]{0.776,0.878,0.706}}Fog      & {\cellcolor[rgb]{0.706,0.776,0.906}}Cloud    & {\cellcolor[rgb]{0.776,0.878,0.706}}Fog      & {\cellcolor[rgb]{0.706,0.776,0.906}}Cloud    & {\cellcolor[rgb]{0.776,0.878,0.706}}Fog      & {\cellcolor[rgb]{0.706,0.776,0.906}}Cloud     \\ 
\hline
\multirow{8}{*}{CNN-AE} & {\cellcolor[rgb]{0.557,0.663,0.859}}Average  & {\cellcolor[rgb]{0.957,0.69,0.518}}TF                 & {\cellcolor[rgb]{0.776,0.878,0.706}}1.01413  & {\cellcolor[rgb]{0.706,0.776,0.906}}23.3021  & {\cellcolor[rgb]{0.776,0.878,0.706}}0.418136 & {\cellcolor[rgb]{0.706,0.776,0.906}}0.418136 & {\cellcolor[rgb]{0.776,0.878,0.706}}0.323962 & {\cellcolor[rgb]{0.706,0.776,0.906}}0.323962 & {\cellcolor[rgb]{0.776,0.878,0.706}}0.311666 & {\cellcolor[rgb]{0.706,0.776,0.906}}0.311666 & {\cellcolor[rgb]{0.776,0.878,0.706}}0.91451  & {\cellcolor[rgb]{0.706,0.776,0.906}}0.91451  & {\cellcolor[rgb]{0.776,0.878,0.706}}0.464895 & {\cellcolor[rgb]{0.706,0.776,0.906}}0.464895 & {\cellcolor[rgb]{0.776,0.878,0.706}}6.88913  & {\cellcolor[rgb]{0.706,0.776,0.906}}1.416631 & {\cellcolor[rgb]{0.776,0.878,0.706}}34.64844 & {\cellcolor[rgb]{0.706,0.776,0.906}}291.4189  \\
                        & {\cellcolor[rgb]{0.557,0.663,0.859}}Kurtosis & {\cellcolor[rgb]{0.957,0.69,0.518}}TF                 & {\cellcolor[rgb]{0.776,0.878,0.706}}0.995147 & {\cellcolor[rgb]{0.706,0.776,0.906}}24.0796  & {\cellcolor[rgb]{0.776,0.878,0.706}}0.519019 & {\cellcolor[rgb]{0.706,0.776,0.906}}0.519019 & {\cellcolor[rgb]{0.776,0.878,0.706}}0.728089 & {\cellcolor[rgb]{0.706,0.776,0.906}}0.728089 & {\cellcolor[rgb]{0.776,0.878,0.706}}0.755389 & {\cellcolor[rgb]{0.706,0.776,0.906}}0.755389 & {\cellcolor[rgb]{0.776,0.878,0.706}}0.945001 & {\cellcolor[rgb]{0.706,0.776,0.906}}0.945001 & {\cellcolor[rgb]{0.776,0.878,0.706}}0.839623 & {\cellcolor[rgb]{0.706,0.776,0.906}}0.839623 & {\cellcolor[rgb]{0.776,0.878,0.706}}67.18984 & {\cellcolor[rgb]{0.706,0.776,0.906}}23.71106 & {\cellcolor[rgb]{0.776,0.878,0.706}}34.60938 & {\cellcolor[rgb]{0.706,0.776,0.906}}291.0166  \\
                        & {\cellcolor[rgb]{0.557,0.663,0.859}}MAD      & {\cellcolor[rgb]{0.957,0.69,0.518}}TF                 & {\cellcolor[rgb]{0.776,0.878,0.706}}1.025978 & {\cellcolor[rgb]{0.706,0.776,0.906}}24.03    & {\cellcolor[rgb]{0.776,0.878,0.706}}0.448848 & {\cellcolor[rgb]{0.706,0.776,0.906}}0.448848 & {\cellcolor[rgb]{0.776,0.878,0.706}}0.7072   & {\cellcolor[rgb]{0.706,0.776,0.906}}0.7072   & {\cellcolor[rgb]{0.776,0.878,0.706}}0.740935 & {\cellcolor[rgb]{0.706,0.776,0.906}}0.740935 & {\cellcolor[rgb]{0.776,0.878,0.706}}0.934799 & {\cellcolor[rgb]{0.706,0.776,0.906}}0.934799 & {\cellcolor[rgb]{0.776,0.878,0.706}}0.826653 & {\cellcolor[rgb]{0.706,0.776,0.906}}0.826653 & {\cellcolor[rgb]{0.776,0.878,0.706}}72.07326 & {\cellcolor[rgb]{0.706,0.776,0.906}}21.63906 & {\cellcolor[rgb]{0.776,0.878,0.706}}34.64844 & {\cellcolor[rgb]{0.706,0.776,0.906}}291.4189  \\
                        & {\cellcolor[rgb]{1,0.851,0.4}}MM             & {\cellcolor[rgb]{0.957,0.69,0.518}}TF                 & {\cellcolor[rgb]{0.776,0.878,0.706}}440.2933 & {\cellcolor[rgb]{0.706,0.776,0.906}}96.6556  & {\cellcolor[rgb]{0.776,0.878,0.706}}0.498538 & {\cellcolor[rgb]{0.706,0.776,0.906}}0.498538 & {\cellcolor[rgb]{0.776,0.878,0.706}}0.18035  & {\cellcolor[rgb]{0.706,0.776,0.906}}0.18035  & {\cellcolor[rgb]{0.776,0.878,0.706}}0.138802 & {\cellcolor[rgb]{0.706,0.776,0.906}}0.138802 & {\cellcolor[rgb]{0.776,0.878,0.706}}0.941108 & {\cellcolor[rgb]{0.706,0.776,0.906}}0.941108 & {\cellcolor[rgb]{0.776,0.878,0.706}}0.241923 & {\cellcolor[rgb]{0.706,0.776,0.906}}0.241923 & {\cellcolor[rgb]{0.776,0.878,0.706}}1.433161 & {\cellcolor[rgb]{0.706,0.776,0.906}}0.481541 & {\cellcolor[rgb]{0.776,0.878,0.706}}397203.7 & {\cellcolor[rgb]{0.706,0.776,0.906}}1191798   \\
                        & {\cellcolor[rgb]{1,0.851,0.4}}NS             & {\cellcolor[rgb]{0.957,0.69,0.518}}TF                 & {\cellcolor[rgb]{0.776,0.878,0.706}}440      & {\cellcolor[rgb]{0.706,0.776,0.906}}103.3094 & {\cellcolor[rgb]{0.776,0.878,0.706}}0.5      & {\cellcolor[rgb]{0.706,0.776,0.906}}0.50022  & {\cellcolor[rgb]{0.776,0.878,0.706}}0.059669 & {\cellcolor[rgb]{0.706,0.776,0.906}}0.059669 & {\cellcolor[rgb]{0.776,0.878,0.706}}0.002144        & {\cellcolor[rgb]{0.706,0.776,0.906}}0.002144 & {\cellcolor[rgb]{0.776,0.878,0.706}}0.942253 & {\cellcolor[rgb]{0.706,0.776,0.906}}0.953552 & {\cellcolor[rgb]{0.776,0.878,0.706}}0.004278 & {\cellcolor[rgb]{0.706,0.776,0.906}}0.004278 & {\cellcolor[rgb]{0.776,0.878,0.706}}NA       & {\cellcolor[rgb]{0.706,0.776,0.906}}NA       & {\cellcolor[rgb]{0.776,0.878,0.706}}397194.2 & {\cellcolor[rgb]{0.706,0.776,0.906}}1191799   \\
                        & {\cellcolor[rgb]{0.557,0.663,0.859}}Skew     & {\cellcolor[rgb]{0.957,0.69,0.518}}TF                 & {\cellcolor[rgb]{0.776,0.878,0.706}}1.001088 & {\cellcolor[rgb]{0.706,0.776,0.906}}24.6464  & {\cellcolor[rgb]{0.776,0.878,0.706}}0.361076 & {\cellcolor[rgb]{0.706,0.776,0.906}}0.361073 & {\cellcolor[rgb]{0.776,0.878,0.706}}0.261203 & {\cellcolor[rgb]{0.706,0.776,0.906}}0.261197 & {\cellcolor[rgb]{0.776,0.878,0.706}}0.248162 & {\cellcolor[rgb]{0.706,0.776,0.906}}0.248156 & {\cellcolor[rgb]{0.776,0.878,0.706}}0.885031 & {\cellcolor[rgb]{0.706,0.776,0.906}}0.885028 & {\cellcolor[rgb]{0.776,0.878,0.706}}0.387632 & {\cellcolor[rgb]{0.706,0.776,0.906}}0.387624 & {\cellcolor[rgb]{0.776,0.878,0.706}}69.34095 & {\cellcolor[rgb]{0.706,0.776,0.906}}34.05004 & {\cellcolor[rgb]{0.776,0.878,0.706}}34.60938 & {\cellcolor[rgb]{0.706,0.776,0.906}}291.0166  \\
                        & {\cellcolor[rgb]{1,0.851,0.4}}SS             & {\cellcolor[rgb]{0.957,0.69,0.518}}TF                 & {\cellcolor[rgb]{0.776,0.878,0.706}}440      & {\cellcolor[rgb]{0.706,0.776,0.906}}106.2    & {\cellcolor[rgb]{0.776,0.878,0.706}}0.699918 & {\cellcolor[rgb]{0.706,0.776,0.906}}0.699918 & {\cellcolor[rgb]{0.776,0.878,0.706}}0.758349 & {\cellcolor[rgb]{0.706,0.776,0.906}}0.758349 & {\cellcolor[rgb]{0.776,0.878,0.706}}0.765979 & {\cellcolor[rgb]{0.706,0.776,0.906}}0.765979 & {\cellcolor[rgb]{0.776,0.878,0.706}}0.971539 & {\cellcolor[rgb]{0.706,0.776,0.906}}0.971539 & {\cellcolor[rgb]{0.776,0.878,0.706}}0.856599 & {\cellcolor[rgb]{0.706,0.776,0.906}}0.856599 & {\cellcolor[rgb]{0.776,0.878,0.706}}0.814415 & {\cellcolor[rgb]{0.706,0.776,0.906}}0.790764 & {\cellcolor[rgb]{0.776,0.878,0.706}}397203.6 & {\cellcolor[rgb]{0.706,0.776,0.906}}1191796   \\
                        & {\cellcolor[rgb]{0.557,0.663,0.859}}StDev    & {\cellcolor[rgb]{0.957,0.69,0.518}}TF                 & {\cellcolor[rgb]{0.776,0.878,0.706}}1.051563 & {\cellcolor[rgb]{0.706,0.776,0.906}}25.3799  & {\cellcolor[rgb]{0.776,0.878,0.706}}0.497397 & {\cellcolor[rgb]{0.706,0.776,0.906}}0.497397 & {\cellcolor[rgb]{0.776,0.878,0.706}}0.929901 & {\cellcolor[rgb]{0.706,0.776,0.906}}0.929901 & {\cellcolor[rgb]{0.776,0.878,0.706}}0.986375 & {\cellcolor[rgb]{0.706,0.776,0.906}}0.986375 & {\cellcolor[rgb]{0.776,0.878,0.706}}0.941966 & {\cellcolor[rgb]{0.706,0.776,0.906}}0.941966 & {\cellcolor[rgb]{0.776,0.878,0.706}}0.963659 & {\cellcolor[rgb]{0.706,0.776,0.906}}0.963659 & {\cellcolor[rgb]{0.776,0.878,0.706}}20.04371 & {\cellcolor[rgb]{0.706,0.776,0.906}}3.039728 & {\cellcolor[rgb]{0.776,0.878,0.706}}34.64844 & {\cellcolor[rgb]{0.706,0.776,0.906}}291.4189  \\ 
\hline
\multirow{5}{*}{RNN}    & {\cellcolor[rgb]{0.557,0.663,0.859}}Average  & {\cellcolor[rgb]{0.957,0.69,0.518}}TF                 & {\cellcolor[rgb]{0.776,0.878,0.706}}8.164231 & {\cellcolor[rgb]{0.706,0.776,0.906}}29.9278  & {\cellcolor[rgb]{0.776,0.878,0.706}}0.431887 & {\cellcolor[rgb]{0.706,0.776,0.906}}0.431887 & {\cellcolor[rgb]{0.776,0.878,0.706}}0.174428 & {\cellcolor[rgb]{0.706,0.776,0.906}}0.174428 & {\cellcolor[rgb]{0.776,0.878,0.706}}0.140811 & {\cellcolor[rgb]{0.706,0.776,0.906}}0.140811 & {\cellcolor[rgb]{0.776,0.878,0.706}}0.892397 & {\cellcolor[rgb]{0.706,0.776,0.906}}0.892397 & {\cellcolor[rgb]{0.776,0.878,0.706}}0.243241 & {\cellcolor[rgb]{0.706,0.776,0.906}}0.243241 & {\cellcolor[rgb]{0.776,0.878,0.706}}6.911079 & {\cellcolor[rgb]{0.706,0.776,0.906}}1.273132 & {\cellcolor[rgb]{0.776,0.878,0.706}}797.8203 & {\cellcolor[rgb]{0.706,0.776,0.906}}4045.537  \\
                        & {\cellcolor[rgb]{0.557,0.663,0.859}}Kurtosis & {\cellcolor[rgb]{0.957,0.69,0.518}}TF                 & {\cellcolor[rgb]{0.776,0.878,0.706}}8.020618 & {\cellcolor[rgb]{0.706,0.776,0.906}}30.07291 & {\cellcolor[rgb]{0.776,0.878,0.706}}0.515931 & {\cellcolor[rgb]{0.706,0.776,0.906}}0.515931 & {\cellcolor[rgb]{0.776,0.878,0.706}}0.433345 & {\cellcolor[rgb]{0.706,0.776,0.906}}0.433345 & {\cellcolor[rgb]{0.776,0.878,0.706}}0.422561 & {\cellcolor[rgb]{0.706,0.776,0.906}}0.422561 & {\cellcolor[rgb]{0.776,0.878,0.706}}0.946374 & {\cellcolor[rgb]{0.706,0.776,0.906}}0.946374 & {\cellcolor[rgb]{0.776,0.878,0.706}}0.584251 & {\cellcolor[rgb]{0.706,0.776,0.906}}0.584251 & {\cellcolor[rgb]{0.776,0.878,0.706}}67.16076 & {\cellcolor[rgb]{0.706,0.776,0.906}}22.99305 & {\cellcolor[rgb]{0.776,0.878,0.706}}797.8047 & {\cellcolor[rgb]{0.706,0.776,0.906}}4042.888  \\
                        & {\cellcolor[rgb]{0.557,0.663,0.859}}MAD      & {\cellcolor[rgb]{0.957,0.69,0.518}}TF                 & {\cellcolor[rgb]{0.776,0.878,0.706}}8.014902 & {\cellcolor[rgb]{0.706,0.776,0.906}}29.02    & {\cellcolor[rgb]{0.776,0.878,0.706}}0.421892 & {\cellcolor[rgb]{0.706,0.776,0.906}}0.421892 & {\cellcolor[rgb]{0.776,0.878,0.706}}0.69089  & {\cellcolor[rgb]{0.706,0.776,0.906}}0.69089  & {\cellcolor[rgb]{0.776,0.878,0.706}}0.726014 & {\cellcolor[rgb]{0.706,0.776,0.906}}0.726014 & {\cellcolor[rgb]{0.776,0.878,0.706}}0.930689 & {\cellcolor[rgb]{0.706,0.776,0.906}}0.930689 & {\cellcolor[rgb]{0.776,0.878,0.706}}0.815708 & {\cellcolor[rgb]{0.706,0.776,0.906}}0.815708 & {\cellcolor[rgb]{0.776,0.878,0.706}}71.88432 & {\cellcolor[rgb]{0.706,0.776,0.906}}19.5469  & {\cellcolor[rgb]{0.776,0.878,0.706}}797.8164 & {\cellcolor[rgb]{0.706,0.776,0.906}}4044.854  \\
                        & {\cellcolor[rgb]{0.557,0.663,0.859}}Skew     & {\cellcolor[rgb]{0.957,0.69,0.518}}TF                 & {\cellcolor[rgb]{0.776,0.878,0.706}}7.621182 & {\cellcolor[rgb]{0.706,0.776,0.906}}29.82979 & {\cellcolor[rgb]{0.776,0.878,0.706}}0.448924 & {\cellcolor[rgb]{0.706,0.776,0.906}}0.448965 & {\cellcolor[rgb]{0.776,0.878,0.706}}0.544221 & {\cellcolor[rgb]{0.706,0.776,0.906}}0.544209 & {\cellcolor[rgb]{0.776,0.878,0.706}}0.556664 & {\cellcolor[rgb]{0.706,0.776,0.906}}0.556646 & {\cellcolor[rgb]{0.776,0.878,0.706}}0.932372 & {\cellcolor[rgb]{0.706,0.776,0.906}}0.93238  & {\cellcolor[rgb]{0.776,0.878,0.706}}0.697119 & {\cellcolor[rgb]{0.706,0.776,0.906}}0.697107 & {\cellcolor[rgb]{0.776,0.878,0.706}}69.21164 & {\cellcolor[rgb]{0.706,0.776,0.906}}31.53431 & {\cellcolor[rgb]{0.776,0.878,0.706}}797.8125 & {\cellcolor[rgb]{0.706,0.776,0.906}}4044.202  \\
                        & {\cellcolor[rgb]{0.557,0.663,0.859}}StDev    & {\cellcolor[rgb]{0.957,0.69,0.518}}TF                 & {\cellcolor[rgb]{0.776,0.878,0.706}}8.09226  & {\cellcolor[rgb]{0.706,0.776,0.906}}30.83805 & {\cellcolor[rgb]{0.776,0.878,0.706}}0.420087 & {\cellcolor[rgb]{0.706,0.776,0.906}}0.42009  & {\cellcolor[rgb]{0.776,0.878,0.706}}0.436152 & {\cellcolor[rgb]{0.706,0.776,0.906}}0.436158 & {\cellcolor[rgb]{0.776,0.878,0.706}}0.43825  & {\cellcolor[rgb]{0.706,0.776,0.906}}0.438256 & {\cellcolor[rgb]{0.776,0.878,0.706}}0.922818 & {\cellcolor[rgb]{0.706,0.776,0.906}}0.922819 & {\cellcolor[rgb]{0.776,0.878,0.706}}0.594276 & {\cellcolor[rgb]{0.706,0.776,0.906}}0.594282 & {\cellcolor[rgb]{0.776,0.878,0.706}}20.04073 & {\cellcolor[rgb]{0.706,0.776,0.906}}3.640271 & {\cellcolor[rgb]{0.776,0.878,0.706}}797.8203 & {\cellcolor[rgb]{0.706,0.776,0.906}}4045.537  \\ 
\hline
\multirow{3}{*}{RNN-AE} & {\cellcolor[rgb]{1,0.851,0.4}}MM             & {\cellcolor[rgb]{0.957,0.69,0.518}}TF                 & {\cellcolor[rgb]{0.776,0.878,0.706}}18.32667 & {\cellcolor[rgb]{0.706,0.776,0.906}}40.9852  & {\cellcolor[rgb]{0.776,0.878,0.706}}0.470241 & {\cellcolor[rgb]{0.706,0.776,0.906}}0.470231 & {\cellcolor[rgb]{0.776,0.878,0.706}}0.166829 & {\cellcolor[rgb]{0.706,0.776,0.906}}0.166811 & {\cellcolor[rgb]{0.776,0.878,0.706}}0.127211 & {\cellcolor[rgb]{0.706,0.776,0.906}}0.127192 & {\cellcolor[rgb]{0.776,0.878,0.706}}0.917464 & {\cellcolor[rgb]{0.706,0.776,0.906}}0.917453 & {\cellcolor[rgb]{0.776,0.878,0.706}}0.22344  & {\cellcolor[rgb]{0.706,0.776,0.906}}0.223411 & {\cellcolor[rgb]{0.776,0.878,0.706}}1.330966 & {\cellcolor[rgb]{0.706,0.776,0.906}}0.328342 & {\cellcolor[rgb]{0.776,0.878,0.706}}1414.535 & {\cellcolor[rgb]{0.706,0.776,0.906}}10386.43  \\
                        & {\cellcolor[rgb]{1,0.851,0.4}}NS             & {\cellcolor[rgb]{0.957,0.69,0.518}}TF                 & {\cellcolor[rgb]{0.776,0.878,0.706}}17.90922 & {\cellcolor[rgb]{0.706,0.776,0.906}}41.1471  & {\cellcolor[rgb]{0.776,0.878,0.706}}0.5      & {\cellcolor[rgb]{0.706,0.776,0.906}}0.500252 & {\cellcolor[rgb]{0.776,0.878,0.706}}0.942253 & {\cellcolor[rgb]{0.706,0.776,0.906}}0.05964  & {\cellcolor[rgb]{0.776,0.878,0.706}}1        & {\cellcolor[rgb]{0.706,0.776,0.906}}0.002107 & {\cellcolor[rgb]{0.776,0.878,0.706}}0.942253 & {\cellcolor[rgb]{0.706,0.776,0.906}}0.955432 & {\cellcolor[rgb]{0.776,0.878,0.706}}0.970268 & {\cellcolor[rgb]{0.706,0.776,0.906}}0.004205 & {\cellcolor[rgb]{0.776,0.878,0.706}}NA       & {\cellcolor[rgb]{0.706,0.776,0.906}}NA       & {\cellcolor[rgb]{0.776,0.878,0.706}}1410.465 & {\cellcolor[rgb]{0.706,0.776,0.906}}10386.67  \\
                        & {\cellcolor[rgb]{1,0.851,0.4}}SS             & {\cellcolor[rgb]{0.957,0.69,0.518}}TF                 & {\cellcolor[rgb]{0.776,0.878,0.706}}18.08283 & {\cellcolor[rgb]{0.706,0.776,0.906}}49.55683 & {\cellcolor[rgb]{0.776,0.878,0.706}}0.699284 & {\cellcolor[rgb]{0.706,0.776,0.906}}0.699284 & {\cellcolor[rgb]{0.776,0.878,0.706}}0.671158 & {\cellcolor[rgb]{0.706,0.776,0.906}}0.671158 & {\cellcolor[rgb]{0.776,0.878,0.706}}0.667486 & {\cellcolor[rgb]{0.706,0.776,0.906}}0.667486 & {\cellcolor[rgb]{0.776,0.878,0.706}}0.975904 & {\cellcolor[rgb]{0.706,0.776,0.906}}0.975904 & {\cellcolor[rgb]{0.776,0.878,0.706}}0.792754 & {\cellcolor[rgb]{0.706,0.776,0.906}}0.792754 & {\cellcolor[rgb]{0.776,0.878,0.706}}3.304777 & {\cellcolor[rgb]{0.706,0.776,0.906}}1.005414 & {\cellcolor[rgb]{0.776,0.878,0.706}}1414.512 & {\cellcolor[rgb]{0.706,0.776,0.906}}10378.48  \\ 
\hline
\multirow{8}{*}{OC-SVM} & {\cellcolor[rgb]{0.557,0.663,0.859}}Average  & {\cellcolor[rgb]{0.502,0,0.502}}\textcolor{white}{SK} & {\cellcolor[rgb]{0.776,0.878,0.706}}0.894195 & {\cellcolor[rgb]{0.706,0.776,0.906}}0.342938 & {\cellcolor[rgb]{0.776,0.878,0.706}}0.491203 & {\cellcolor[rgb]{0.706,0.776,0.906}}0.491203 & {\cellcolor[rgb]{0.776,0.878,0.706}}0.7086   & {\cellcolor[rgb]{0.706,0.776,0.906}}0.7086   & {\cellcolor[rgb]{0.776,0.878,0.706}}0.801859 & {\cellcolor[rgb]{0.706,0.776,0.906}}0.801859 & {\cellcolor[rgb]{0.776,0.878,0.706}}0.84711  & {\cellcolor[rgb]{0.706,0.776,0.906}}0.84711  & {\cellcolor[rgb]{0.776,0.878,0.706}}0.823864 & {\cellcolor[rgb]{0.706,0.776,0.906}}0.823864 & {\cellcolor[rgb]{0.776,0.878,0.706}}6.88913  & {\cellcolor[rgb]{0.706,0.776,0.906}}1.416631 & {\cellcolor[rgb]{0.776,0.878,0.706}}56.62305 & {\cellcolor[rgb]{0.706,0.776,0.906}}56.62305  \\
                        & {\cellcolor[rgb]{0.557,0.663,0.859}}Kurtosis & {\cellcolor[rgb]{0.502,0,0.502}}\textcolor{white}{SK} & {\cellcolor[rgb]{0.776,0.878,0.706}}0.855004 & {\cellcolor[rgb]{0.706,0.776,0.906}}0.341614 & {\cellcolor[rgb]{0.776,0.878,0.706}}0.481406 & {\cellcolor[rgb]{0.706,0.776,0.906}}0.481406 & {\cellcolor[rgb]{0.776,0.878,0.706}}0.8127   & {\cellcolor[rgb]{0.706,0.776,0.906}}0.8127   & {\cellcolor[rgb]{0.776,0.878,0.706}}0.954818 & {\cellcolor[rgb]{0.706,0.776,0.906}}0.954818 & {\cellcolor[rgb]{0.776,0.878,0.706}}0.84496  & {\cellcolor[rgb]{0.706,0.776,0.906}}0.84496  & {\cellcolor[rgb]{0.776,0.878,0.706}}0.896536 & {\cellcolor[rgb]{0.706,0.776,0.906}}0.896536 & {\cellcolor[rgb]{0.776,0.878,0.706}}67.18984 & {\cellcolor[rgb]{0.706,0.776,0.906}}23.71106 & {\cellcolor[rgb]{0.776,0.878,0.706}}50.1582  & {\cellcolor[rgb]{0.706,0.776,0.906}}50.1582   \\
                        & {\cellcolor[rgb]{0.557,0.663,0.859}}MAD      & {\cellcolor[rgb]{0.502,0,0.502}}\textcolor{white}{SK} & {\cellcolor[rgb]{0.776,0.878,0.706}}0.512389 & {\cellcolor[rgb]{0.706,0.776,0.906}}0.207312 & {\cellcolor[rgb]{0.776,0.878,0.706}}0.500747 & {\cellcolor[rgb]{0.706,0.776,0.906}}0.500747 & {\cellcolor[rgb]{0.776,0.878,0.706}}0.7141   & {\cellcolor[rgb]{0.706,0.776,0.906}}0.7141   & {\cellcolor[rgb]{0.776,0.878,0.706}}0.805624 & {\cellcolor[rgb]{0.706,0.776,0.906}}0.805624 & {\cellcolor[rgb]{0.776,0.878,0.706}}0.850137 & {\cellcolor[rgb]{0.706,0.776,0.906}}0.850137 & {\cellcolor[rgb]{0.776,0.878,0.706}}0.827282 & {\cellcolor[rgb]{0.706,0.776,0.906}}0.827282 & {\cellcolor[rgb]{0.776,0.878,0.706}}72.07326 & {\cellcolor[rgb]{0.706,0.776,0.906}}21.63906 & {\cellcolor[rgb]{0.776,0.878,0.706}}8.673828 & {\cellcolor[rgb]{0.706,0.776,0.906}}8.673828  \\
                        & {\cellcolor[rgb]{1,0.851,0.4}}MM             & {\cellcolor[rgb]{0.502,0,0.502}}\textcolor{white}{SK} & {\cellcolor[rgb]{0.776,0.878,0.706}}0.687799 & {\cellcolor[rgb]{0.706,0.776,0.906}}0.367881 & {\cellcolor[rgb]{0.776,0.878,0.706}}0.5      & {\cellcolor[rgb]{0.706,0.776,0.906}}0.5      & {\cellcolor[rgb]{0.776,0.878,0.706}}0.1501   & {\cellcolor[rgb]{0.706,0.776,0.906}}0.1501   & {\cellcolor[rgb]{0.776,0.878,0.706}}0        & {\cellcolor[rgb]{0.706,0.776,0.906}}0        & {\cellcolor[rgb]{0.776,0.878,0.706}}1        & {\cellcolor[rgb]{0.706,0.776,0.906}}1        & {\cellcolor[rgb]{0.776,0.878,0.706}}0        & {\cellcolor[rgb]{0.706,0.776,0.906}}0        & {\cellcolor[rgb]{0.776,0.878,0.706}}1.433161 & {\cellcolor[rgb]{0.706,0.776,0.906}}0.481541 & {\cellcolor[rgb]{0.776,0.878,0.706}}407.5889 & {\cellcolor[rgb]{0.706,0.776,0.906}}407.5889  \\
                        & {\cellcolor[rgb]{1,0.851,0.4}}NS             & {\cellcolor[rgb]{0.502,0,0.502}}\textcolor{white}{SK} & {\cellcolor[rgb]{0.776,0.878,0.706}}0.680684 & {\cellcolor[rgb]{0.706,0.776,0.906}}0.335778 & {\cellcolor[rgb]{0.776,0.878,0.706}}0.5      & {\cellcolor[rgb]{0.706,0.776,0.906}}0.5      & {\cellcolor[rgb]{0.776,0.878,0.706}}0.1501   & {\cellcolor[rgb]{0.706,0.776,0.906}}0.1501   & {\cellcolor[rgb]{0.776,0.878,0.706}}0        & {\cellcolor[rgb]{0.706,0.776,0.906}}0        & {\cellcolor[rgb]{0.776,0.878,0.706}}1        & {\cellcolor[rgb]{0.706,0.776,0.906}}1        & {\cellcolor[rgb]{0.776,0.878,0.706}}0        & {\cellcolor[rgb]{0.706,0.776,0.906}}0        & {\cellcolor[rgb]{0.776,0.878,0.706}}NA       & {\cellcolor[rgb]{0.706,0.776,0.906}}NA       & {\cellcolor[rgb]{0.776,0.878,0.706}}395.542  & {\cellcolor[rgb]{0.706,0.776,0.906}}395.542   \\
                        & {\cellcolor[rgb]{1,0.851,0.4}}SS             & {\cellcolor[rgb]{0.502,0,0.502}}\textcolor{white}{SK} & {\cellcolor[rgb]{0.776,0.878,0.706}}1.314422 & {\cellcolor[rgb]{0.706,0.776,0.906}}0.576975 & {\cellcolor[rgb]{0.776,0.878,0.706}}0.496702 & {\cellcolor[rgb]{0.706,0.776,0.906}}0.496702 & {\cellcolor[rgb]{0.776,0.878,0.706}}0.8028   & {\cellcolor[rgb]{0.706,0.776,0.906}}0.8028   & {\cellcolor[rgb]{0.776,0.878,0.706}}0.93411  & {\cellcolor[rgb]{0.706,0.776,0.906}}0.93411  & {\cellcolor[rgb]{0.776,0.878,0.706}}0.849    & {\cellcolor[rgb]{0.706,0.776,0.906}}0.849    & {\cellcolor[rgb]{0.776,0.878,0.706}}0.889524 & {\cellcolor[rgb]{0.706,0.776,0.906}}0.889524 & {\cellcolor[rgb]{0.776,0.878,0.706}}0.814415 & {\cellcolor[rgb]{0.706,0.776,0.906}}0.790764 & {\cellcolor[rgb]{0.776,0.878,0.706}}1442.616 & {\cellcolor[rgb]{0.706,0.776,0.906}}1442.616  \\
                        & {\cellcolor[rgb]{0.557,0.663,0.859}}Skew     & {\cellcolor[rgb]{0.502,0,0.502}}\textcolor{white}{SK} & {\cellcolor[rgb]{0.776,0.878,0.706}}0.545252 & {\cellcolor[rgb]{0.706,0.776,0.906}}0.189985 & {\cellcolor[rgb]{0.776,0.878,0.706}}0.493176 & {\cellcolor[rgb]{0.706,0.776,0.906}}0.493176 & {\cellcolor[rgb]{0.776,0.878,0.706}}0.8383   & {\cellcolor[rgb]{0.706,0.776,0.906}}0.8383   & {\cellcolor[rgb]{0.776,0.878,0.706}}0.986351 & {\cellcolor[rgb]{0.706,0.776,0.906}}0.986351 & {\cellcolor[rgb]{0.776,0.878,0.706}}0.848138 & {\cellcolor[rgb]{0.706,0.776,0.906}}0.848138 & {\cellcolor[rgb]{0.776,0.878,0.706}}0.912038 & {\cellcolor[rgb]{0.706,0.776,0.906}}0.912038 & {\cellcolor[rgb]{0.776,0.878,0.706}}69.34095 & {\cellcolor[rgb]{0.706,0.776,0.906}}34.05004 & {\cellcolor[rgb]{0.776,0.878,0.706}}10.2168  & {\cellcolor[rgb]{0.706,0.776,0.906}}10.2168   \\
                        & {\cellcolor[rgb]{0.557,0.663,0.859}}StDev    & {\cellcolor[rgb]{0.502,0,0.502}}\textcolor{white}{SK} & {\cellcolor[rgb]{0.776,0.878,0.706}}1.13791  & {\cellcolor[rgb]{0.706,0.776,0.906}}0.469978 & {\cellcolor[rgb]{0.776,0.878,0.706}}0.501338 & {\cellcolor[rgb]{0.706,0.776,0.906}}0.501338 & {\cellcolor[rgb]{0.776,0.878,0.706}}0.7897   & {\cellcolor[rgb]{0.706,0.776,0.906}}0.7897   & {\cellcolor[rgb]{0.776,0.878,0.706}}0.913402 & {\cellcolor[rgb]{0.706,0.776,0.906}}0.913402 & {\cellcolor[rgb]{0.776,0.878,0.706}}0.850274 & {\cellcolor[rgb]{0.706,0.776,0.906}}0.850274 & {\cellcolor[rgb]{0.776,0.878,0.706}}0.880708 & {\cellcolor[rgb]{0.706,0.776,0.906}}0.880708 & {\cellcolor[rgb]{0.776,0.878,0.706}}20.04371 & {\cellcolor[rgb]{0.706,0.776,0.906}}3.039728 & {\cellcolor[rgb]{0.776,0.878,0.706}}92.62793 & {\cellcolor[rgb]{0.706,0.776,0.906}}92.62793  \\ 
\hline
\multirow{8}{*}{IF}     & {\cellcolor[rgb]{0.557,0.663,0.859}}Average  & {\cellcolor[rgb]{0.502,0,0.502}}\textcolor{white}{SK} & {\cellcolor[rgb]{0.776,0.878,0.706}}0.1884   & {\cellcolor[rgb]{0.706,0.776,0.906}}0.2868   & {\cellcolor[rgb]{0.776,0.878,0.706}}0.5891   & {\cellcolor[rgb]{0.706,0.776,0.906}}0.5891   & {\cellcolor[rgb]{0.776,0.878,0.706}}0.7152   & {\cellcolor[rgb]{0.706,0.776,0.906}}0.7152   & {\cellcolor[rgb]{0.776,0.878,0.706}}0.7179~  & {\cellcolor[rgb]{0.706,0.776,0.906}}0.7179   & {\cellcolor[rgb]{0.776,0.878,0.706}}0.9921   & {\cellcolor[rgb]{0.706,0.776,0.906}}0.9921   & {\cellcolor[rgb]{0.776,0.878,0.706}}0.8330   & {\cellcolor[rgb]{0.706,0.776,0.906}}0.8330   & {\cellcolor[rgb]{0.776,0.878,0.706}}6.88913  & {\cellcolor[rgb]{0.706,0.776,0.906}}1.416631 & {\cellcolor[rgb]{0.776,0.878,0.706}}962.10       & {\cellcolor[rgb]{0.706,0.776,0.906}}962.10        \\
                        & {\cellcolor[rgb]{0.557,0.663,0.859}}Kurtosis & {\cellcolor[rgb]{0.502,0,0.502}}\textcolor{white}{SK} & {\cellcolor[rgb]{0.776,0.878,0.706}}0.1884   & {\cellcolor[rgb]{0.706,0.776,0.906}}0.2868   & {\cellcolor[rgb]{0.776,0.878,0.706}}0.6433   & {\cellcolor[rgb]{0.706,0.776,0.906}}0.6433   & {\cellcolor[rgb]{0.776,0.878,0.706}}0.7205   & {\cellcolor[rgb]{0.706,0.776,0.906}}0.7205   & {\cellcolor[rgb]{0.776,0.878,0.706}}0.7221~  & {\cellcolor[rgb]{0.706,0.776,0.906}}0.7221   & {\cellcolor[rgb]{0.776,0.878,0.706}}0.9936   & {\cellcolor[rgb]{0.706,0.776,0.906}}0.9936   & {\cellcolor[rgb]{0.776,0.878,0.706}}0.8364   & {\cellcolor[rgb]{0.706,0.776,0.906}}0.8364   & {\cellcolor[rgb]{0.776,0.878,0.706}}67.18984 & {\cellcolor[rgb]{0.706,0.776,0.906}}23.71106 & {\cellcolor[rgb]{0.776,0.878,0.706}}775.20       & {\cellcolor[rgb]{0.706,0.776,0.906}}775.20        \\
                        & {\cellcolor[rgb]{0.557,0.663,0.859}}MAD      & {\cellcolor[rgb]{0.502,0,0.502}}\textcolor{white}{SK} & {\cellcolor[rgb]{0.776,0.878,0.706}}0.1884   & {\cellcolor[rgb]{0.706,0.776,0.906}}0.2868   & {\cellcolor[rgb]{0.776,0.878,0.706}}0.4325   & {\cellcolor[rgb]{0.706,0.776,0.906}}0.4325   & {\cellcolor[rgb]{0.776,0.878,0.706}}0.7123   & {\cellcolor[rgb]{0.706,0.776,0.906}}0.7123   & {\cellcolor[rgb]{0.776,0.878,0.706}}0.7183   & {\cellcolor[rgb]{0.706,0.776,0.906}}0.7183   & {\cellcolor[rgb]{0.776,0.878,0.706}}0.9876   & {\cellcolor[rgb]{0.706,0.776,0.906}}0.9876   & {\cellcolor[rgb]{0.776,0.878,0.706}}0.8317   & {\cellcolor[rgb]{0.706,0.776,0.906}}0.8317   & {\cellcolor[rgb]{0.776,0.878,0.706}}72.07326 & {\cellcolor[rgb]{0.706,0.776,0.906}}21.63906 & {\cellcolor[rgb]{0.776,0.878,0.706}}634.68       & {\cellcolor[rgb]{0.706,0.776,0.906}}634.68        \\
                        & {\cellcolor[rgb]{1,0.851,0.4}}MM             & {\cellcolor[rgb]{0.502,0,0.502}}\textcolor{white}{SK} & {\cellcolor[rgb]{0.776,0.878,0.706}}0.1978   & {\cellcolor[rgb]{0.706,0.776,0.906}}0.3081   & {\cellcolor[rgb]{0.776,0.878,0.706}}0.7155   & {\cellcolor[rgb]{0.706,0.776,0.906}}0.7155   & {\cellcolor[rgb]{0.776,0.878,0.706}}0.8399   & {\cellcolor[rgb]{0.706,0.776,0.906}}0.8399   & {\cellcolor[rgb]{0.776,0.878,0.706}}0.8426   & {\cellcolor[rgb]{0.706,0.776,0.906}}0.8426   & {\cellcolor[rgb]{0.776,0.878,0.706}}0.9948   & {\cellcolor[rgb]{0.706,0.776,0.906}}0.9948   & {\cellcolor[rgb]{0.776,0.878,0.706}}0.9124   & {\cellcolor[rgb]{0.706,0.776,0.906}}0.9124~  & {\cellcolor[rgb]{0.776,0.878,0.706}}1.433161 & {\cellcolor[rgb]{0.706,0.776,0.906}}0.481541 & {\cellcolor[rgb]{0.776,0.878,0.706}}952.58       & {\cellcolor[rgb]{0.706,0.776,0.906}}952.58        \\
                        & {\cellcolor[rgb]{1,0.851,0.4}}NS             & {\cellcolor[rgb]{0.502,0,0.502}}\textcolor{white}{SK} & {\cellcolor[rgb]{0.776,0.878,0.706}}0.1884   & {\cellcolor[rgb]{0.706,0.776,0.906}}0.2868   & {\cellcolor[rgb]{0.776,0.878,0.706}}0.6951   & {\cellcolor[rgb]{0.706,0.776,0.906}}0.6951   & {\cellcolor[rgb]{0.776,0.878,0.706}}0.8444   & {\cellcolor[rgb]{0.706,0.776,0.906}}0.8220   & {\cellcolor[rgb]{0.776,0.878,0.706}}0.8247   & {\cellcolor[rgb]{0.706,0.776,0.906}}0.8247   & {\cellcolor[rgb]{0.776,0.878,0.706}}0.9944   & {\cellcolor[rgb]{0.706,0.776,0.906}}0.9944   & {\cellcolor[rgb]{0.776,0.878,0.706}}0.9016   & {\cellcolor[rgb]{0.706,0.776,0.906}}0.9016   & {\cellcolor[rgb]{0.776,0.878,0.706}}133.64       & {\cellcolor[rgb]{0.706,0.776,0.906}}133.64       & {\cellcolor[rgb]{0.776,0.878,0.706}}NA       & {\cellcolor[rgb]{0.706,0.776,0.906}}NA        \\
                        & {\cellcolor[rgb]{1,0.851,0.4}}SS             & {\cellcolor[rgb]{0.502,0,0.502}}\textcolor{white}{SK} & {\cellcolor[rgb]{0.776,0.878,0.706}}0.2119   & {\cellcolor[rgb]{0.706,0.776,0.906}}0.3103   & {\cellcolor[rgb]{0.776,0.878,0.706}}0.6951   & {\cellcolor[rgb]{0.706,0.776,0.906}}0.6951   & {\cellcolor[rgb]{0.776,0.878,0.706}}0.8444   & {\cellcolor[rgb]{0.706,0.776,0.906}}0.8220   & {\cellcolor[rgb]{0.776,0.878,0.706}}0.8247   & {\cellcolor[rgb]{0.706,0.776,0.906}}0.8247   & {\cellcolor[rgb]{0.776,0.878,0.706}}0.9944   & {\cellcolor[rgb]{0.706,0.776,0.906}}0.9944   & {\cellcolor[rgb]{0.776,0.878,0.706}}0.9016   & {\cellcolor[rgb]{0.706,0.776,0.906}}0.9016   & {\cellcolor[rgb]{0.776,0.878,0.706}}0.814415 & {\cellcolor[rgb]{0.706,0.776,0.906}}0.790764 & {\cellcolor[rgb]{0.776,0.878,0.706}}944.20       & {\cellcolor[rgb]{0.706,0.776,0.906}}944.20        \\
                        & {\cellcolor[rgb]{0.557,0.663,0.859}}Skew     & {\cellcolor[rgb]{0.502,0,0.502}}\textcolor{white}{SK} & {\cellcolor[rgb]{0.776,0.878,0.706}}0.1884   & {\cellcolor[rgb]{0.706,0.776,0.906}}0.2868   & {\cellcolor[rgb]{0.776,0.878,0.706}}0.6415   & {\cellcolor[rgb]{0.706,0.776,0.906}}0.6415   & {\cellcolor[rgb]{0.776,0.878,0.706}}0.7036   & {\cellcolor[rgb]{0.706,0.776,0.906}}0.7036   & {\cellcolor[rgb]{0.776,0.878,0.706}}0.7049   & {\cellcolor[rgb]{0.706,0.776,0.906}}0.7049   & {\cellcolor[rgb]{0.776,0.878,0.706}}0.9937   & {\cellcolor[rgb]{0.706,0.776,0.906}}0.9937   & {\cellcolor[rgb]{0.776,0.878,0.706}}0.8247   & {\cellcolor[rgb]{0.706,0.776,0.906}}0.8247   & {\cellcolor[rgb]{0.776,0.878,0.706}}69.34095 & {\cellcolor[rgb]{0.706,0.776,0.906}}34.05004 & {\cellcolor[rgb]{0.776,0.878,0.706}}761.20       & {\cellcolor[rgb]{0.706,0.776,0.906}}761.20        \\
                        & {\cellcolor[rgb]{0.557,0.663,0.859}}StDev    & {\cellcolor[rgb]{0.502,0,0.502}}\textcolor{white}{SK} & {\cellcolor[rgb]{0.776,0.878,0.706}}0.1884   & {\cellcolor[rgb]{0.706,0.776,0.906}}0.2868   & {\cellcolor[rgb]{0.776,0.878,0.706}}0.5519   & {\cellcolor[rgb]{0.706,0.776,0.906}}0.5519   & {\cellcolor[rgb]{0.776,0.878,0.706}}0.7034   & {\cellcolor[rgb]{0.706,0.776,0.906}}0.7034   & {\cellcolor[rgb]{0.776,0.878,0.706}}0.7066~~ & {\cellcolor[rgb]{0.706,0.776,0.906}}0.7066   & {\cellcolor[rgb]{0.776,0.878,0.706}}0.9910   & {\cellcolor[rgb]{0.706,0.776,0.906}}0.9910   & {\cellcolor[rgb]{0.776,0.878,0.706}}0.8250   & {\cellcolor[rgb]{0.706,0.776,0.906}}0.8250   & {\cellcolor[rgb]{0.776,0.878,0.706}}20.04371 & {\cellcolor[rgb]{0.706,0.776,0.906}}3.039728 & {\cellcolor[rgb]{0.776,0.878,0.706}}765.32       & {\cellcolor[rgb]{0.706,0.776,0.906}}765.32        \\
\bottomrule[1px]
\end{tabular}
\begin{tablenotes}
\setlength\labelsep{0pt}
 \small
 \item NA: Not applied. MM: MinMax scaler. NS: No scaler applied. SS: Standard Scaler. MAD: Median Absolute Deviation. StDev: Standard Deviation. IF: Isolation Forest. AE: Autoencoder. TF: TensorFlow. SK: scikit-learn.
\end{tablenotes}
\vspace{-0.25cm}
\end{threeparttable}
\end{adjustbox}
\end{sidewaystable*}

\begin{table}
\centering
\def\arraystretch{1.2}
\footnotesize
\caption{Evaluation Results of AnoML Dataset}
\label{tab:anomEval}
\begin{adjustbox}{width = \textwidth}
\begin{threeparttable}
\begin{tabular}{lll|ccc|ccc|ccc|ccc|ccc|ccc|cc|ccc} 
\hline
\multicolumn{3}{c|}{Model Details}                                                                           & \multicolumn{3}{c|}{Inference   Time (ms)}                                                                                         & \multicolumn{3}{c|}{Accuracy}                                                                                                     & \multicolumn{3}{c|}{F1   Score}                                                                                                  & \multicolumn{3}{c|}{AUC}                                                                                                           & \multicolumn{3}{c|}{Recall}                                                                                                       & \multicolumn{3}{c|}{Precision}                                                                                                    & \multicolumn{2}{c|}{Scale   Time (s)}                                                 & \multicolumn{3}{c}{Model Size (KB)}                                                                                                \\ 
\hline
Model                & {\cellcolor[rgb]{0.557,0.663,0.859}}SR       & {\cellcolor[rgb]{0.957,0.69,0.518}}API & {\cellcolor[rgb]{0.788,0.788,0.788}}Edge   & {\cellcolor[rgb]{0.776,0.878,0.706}}Fog   & {\cellcolor[rgb]{0.706,0.776,0.906}}Cloud & {\cellcolor[rgb]{0.788,0.788,0.788}}Edge  & {\cellcolor[rgb]{0.776,0.878,0.706}}Fog   & {\cellcolor[rgb]{0.706,0.776,0.906}}Cloud & {\cellcolor[rgb]{0.788,0.788,0.788}}Edge  & {\cellcolor[rgb]{0.776,0.878,0.706}}Fog  & {\cellcolor[rgb]{0.706,0.776,0.906}}Cloud & {\cellcolor[rgb]{0.788,0.788,0.788}}Edge   & {\cellcolor[rgb]{0.776,0.878,0.706}}Fog   & {\cellcolor[rgb]{0.706,0.776,0.906}}Cloud & {\cellcolor[rgb]{0.788,0.788,0.788}}Edge  & {\cellcolor[rgb]{0.776,0.878,0.706}}Fog   & {\cellcolor[rgb]{0.706,0.776,0.906}}Cloud & {\cellcolor[rgb]{0.788,0.788,0.788}}Edge  & {\cellcolor[rgb]{0.776,0.878,0.706}}Fog   & {\cellcolor[rgb]{0.706,0.776,0.906}}Cloud & {\cellcolor[rgb]{0.776,0.878,0.706}}Fog   & {\cellcolor[rgb]{0.706,0.776,0.906}}Cloud & {\cellcolor[rgb]{0.788,0.788,0.788}}Edge  & {\cellcolor[rgb]{0.776,0.878,0.706}}Fog   & {\cellcolor[rgb]{0.706,0.776,0.906}}Cloud  \\ 
\hline
\multirow{5}{*}{CNN} & {\cellcolor[rgb]{0.557,0.663,0.859}}StDev    & {\cellcolor[rgb]{0.957,0.69,0.518}}TF  & {\cellcolor[rgb]{0.788,0.788,0.788}}174.24 & {\cellcolor[rgb]{0.776,0.878,0.706}}1.95  & {\cellcolor[rgb]{0.706,0.776,0.906}}32.22 & {\cellcolor[rgb]{0.788,0.788,0.788}}0.258 & {\cellcolor[rgb]{0.776,0.878,0.706}}0.251 & {\cellcolor[rgb]{0.706,0.776,0.906}}0.251 & {\cellcolor[rgb]{0.788,0.788,0.788}}0.000 & {\cellcolor[rgb]{0.776,0.878,0.706}}0    & {\cellcolor[rgb]{0.706,0.776,0.906}}0     & {\cellcolor[rgb]{0.788,0.788,0.788}}0.4991 & {\cellcolor[rgb]{0.776,0.878,0.706}}0.485 & {\cellcolor[rgb]{0.706,0.776,0.906}}0.485 & {\cellcolor[rgb]{0.788,0.788,0.788}}0.000 & {\cellcolor[rgb]{0.776,0.878,0.706}}0     & {\cellcolor[rgb]{0.706,0.776,0.906}}0     & {\cellcolor[rgb]{0.788,0.788,0.788}}0.000 & {\cellcolor[rgb]{0.776,0.878,0.706}}0     & {\cellcolor[rgb]{0.706,0.776,0.906}}0     & {\cellcolor[rgb]{0.776,0.878,0.706}}0.716 & {\cellcolor[rgb]{0.706,0.776,0.906}}0.12  & {\cellcolor[rgb]{0.788,0.788,0.788}}19.62 & {\cellcolor[rgb]{0.776,0.878,0.706}}3.172 & {\cellcolor[rgb]{0.706,0.776,0.906}}80     \\
                     & {\cellcolor[rgb]{0.557,0.663,0.859}}Average  & {\cellcolor[rgb]{0.957,0.69,0.518}}TF  & {\cellcolor[rgb]{0.788,0.788,0.788}}172.05 & {\cellcolor[rgb]{0.776,0.878,0.706}}1.13  & {\cellcolor[rgb]{0.706,0.776,0.906}}32.63 & {\cellcolor[rgb]{0.788,0.788,0.788}}0.769 & {\cellcolor[rgb]{0.776,0.878,0.706}}0.871 & {\cellcolor[rgb]{0.706,0.776,0.906}}0.871 & {\cellcolor[rgb]{0.788,0.788,0.788}}0.862 & {\cellcolor[rgb]{0.776,0.878,0.706}}0.91 & {\cellcolor[rgb]{0.706,0.776,0.906}}0.91  & {\cellcolor[rgb]{0.788,0.788,0.788}}0.5738 & {\cellcolor[rgb]{0.776,0.878,0.706}}0.867 & {\cellcolor[rgb]{0.706,0.776,0.906}}0.867 & {\cellcolor[rgb]{0.788,0.788,0.788}}0.979 & {\cellcolor[rgb]{0.776,0.878,0.706}}0.876 & {\cellcolor[rgb]{0.706,0.776,0.906}}0.876 & {\cellcolor[rgb]{0.788,0.788,0.788}}0.770 & {\cellcolor[rgb]{0.776,0.878,0.706}}0.946 & {\cellcolor[rgb]{0.706,0.776,0.906}}0.946 & {\cellcolor[rgb]{0.776,0.878,0.706}}0.241 & {\cellcolor[rgb]{0.706,0.776,0.906}}0.04  & {\cellcolor[rgb]{0.788,0.788,0.788}}19.50 & {\cellcolor[rgb]{0.776,0.878,0.706}}3.152 & {\cellcolor[rgb]{0.706,0.776,0.906}}80     \\
                     & {\cellcolor[rgb]{0.557,0.663,0.859}}Skew     & {\cellcolor[rgb]{0.957,0.69,0.518}}TF  & {\cellcolor[rgb]{0.788,0.788,0.788}}NA     & {\cellcolor[rgb]{0.776,0.878,0.706}}8E-15 & {\cellcolor[rgb]{0.706,0.776,0.906}}32.20 & {\cellcolor[rgb]{0.788,0.788,0.788}}NA    & {\cellcolor[rgb]{0.776,0.878,0.706}}0.475 & {\cellcolor[rgb]{0.706,0.776,0.906}}0.475 & {\cellcolor[rgb]{0.788,0.788,0.788}}NA    & {\cellcolor[rgb]{0.776,0.878,0.706}}0.57 & {\cellcolor[rgb]{0.706,0.776,0.906}}0.57  & {\cellcolor[rgb]{0.788,0.788,0.788}}NA     & {\cellcolor[rgb]{0.776,0.878,0.706}}0.474 & {\cellcolor[rgb]{0.706,0.776,0.906}}0.474 & {\cellcolor[rgb]{0.788,0.788,0.788}}NA    & {\cellcolor[rgb]{0.776,0.878,0.706}}0.476 & {\cellcolor[rgb]{0.706,0.776,0.906}}0.476 & {\cellcolor[rgb]{0.788,0.788,0.788}}NA    & {\cellcolor[rgb]{0.776,0.878,0.706}}0.72  & {\cellcolor[rgb]{0.706,0.776,0.906}}0.72  & {\cellcolor[rgb]{0.776,0.878,0.706}}2.487 & {\cellcolor[rgb]{0.706,0.776,0.906}}0.8   & {\cellcolor[rgb]{0.788,0.788,0.788}}19.52 & {\cellcolor[rgb]{0.776,0.878,0.706}}3.156 & {\cellcolor[rgb]{0.706,0.776,0.906}}80     \\
                     & {\cellcolor[rgb]{0.557,0.663,0.859}}Kurtosis & {\cellcolor[rgb]{0.957,0.69,0.518}}TF  & {\cellcolor[rgb]{0.788,0.788,0.788}}NA     & {\cellcolor[rgb]{0.776,0.878,0.706}}2E-07 & {\cellcolor[rgb]{0.706,0.776,0.906}}32.04 & {\cellcolor[rgb]{0.788,0.788,0.788}}NA    & {\cellcolor[rgb]{0.776,0.878,0.706}}0.259 & {\cellcolor[rgb]{0.706,0.776,0.906}}0.259 & {\cellcolor[rgb]{0.788,0.788,0.788}}NA    & {\cellcolor[rgb]{0.776,0.878,0.706}}0    & {\cellcolor[rgb]{0.706,0.776,0.906}}0     & {\cellcolor[rgb]{0.788,0.788,0.788}}NA     & {\cellcolor[rgb]{0.776,0.878,0.706}}0.5   & {\cellcolor[rgb]{0.706,0.776,0.906}}0.5   & {\cellcolor[rgb]{0.788,0.788,0.788}}NA    & {\cellcolor[rgb]{0.776,0.878,0.706}}0     & {\cellcolor[rgb]{0.706,0.776,0.906}}0     & {\cellcolor[rgb]{0.788,0.788,0.788}}NA    & {\cellcolor[rgb]{0.776,0.878,0.706}}1     & {\cellcolor[rgb]{0.706,0.776,0.906}}1     & {\cellcolor[rgb]{0.776,0.878,0.706}}2.425 & {\cellcolor[rgb]{0.706,0.776,0.906}}0.57  & {\cellcolor[rgb]{0.788,0.788,0.788}}19.50 & {\cellcolor[rgb]{0.776,0.878,0.706}}3.152 & {\cellcolor[rgb]{0.706,0.776,0.906}}80     \\
                     & {\cellcolor[rgb]{0.557,0.663,0.859}}MAD      & {\cellcolor[rgb]{0.957,0.69,0.518}}TF  & {\cellcolor[rgb]{0.788,0.788,0.788}}NA     & {\cellcolor[rgb]{0.776,0.878,0.706}}1.81  & {\cellcolor[rgb]{0.706,0.776,0.906}}32.16 & {\cellcolor[rgb]{0.788,0.788,0.788}}NA    & {\cellcolor[rgb]{0.776,0.878,0.706}}0.576 & {\cellcolor[rgb]{0.706,0.776,0.906}}0.576 & {\cellcolor[rgb]{0.788,0.788,0.788}}NA    & {\cellcolor[rgb]{0.776,0.878,0.706}}0.60 & {\cellcolor[rgb]{0.706,0.776,0.906}}0.60  & {\cellcolor[rgb]{0.788,0.788,0.788}}NA     & {\cellcolor[rgb]{0.776,0.878,0.706}}0.702 & {\cellcolor[rgb]{0.706,0.776,0.906}}0.702 & {\cellcolor[rgb]{0.788,0.788,0.788}}NA    & {\cellcolor[rgb]{0.776,0.878,0.706}}0.441 & {\cellcolor[rgb]{0.706,0.776,0.906}}0.441 & {\cellcolor[rgb]{0.788,0.788,0.788}}NA    & {\cellcolor[rgb]{0.776,0.878,0.706}}0.972 & {\cellcolor[rgb]{0.706,0.776,0.906}}0.972 & {\cellcolor[rgb]{0.776,0.878,0.706}}2.558 & {\cellcolor[rgb]{0.706,0.776,0.906}}0.88  & {\cellcolor[rgb]{0.788,0.788,0.788}}19.52 & {\cellcolor[rgb]{0.776,0.878,0.706}}3.156 & {\cellcolor[rgb]{0.706,0.776,0.906}}80     \\
\hline
\end{tabular}
\begin{tablenotes}
\setlength\labelsep{0pt}
 \small
 \item NA: Not applied. MAD: Median Absolute Deviation. StDev: Standard Deviation. IF: Isolation Forest. TF: TensorFlow.
\end{tablenotes}
\vspace{-0.25cm}
\end{threeparttable}
\end{adjustbox}
\end{table}

\section{Lessons Learned}
\label{Sec:lessonslearnt}

IoT infrastructures are expanding thanks to the benefits of ubiquitous computing. The rapid developments in lightweight edge mechanisms made them desirable for labor-intensive tasks such as anomaly detection. The increasing variety of 32-bit microcontrollers allows us to choose a specific device with desired features (e.g., relatively high RAM) that can run ML algorithms. In the edge environment where these microcontrollers are utilized, while ML-based tasks such as keyword spotting, natural language processing, image processing have proven to be promising, anomaly detection is still in a very early phase \cite{banbury2020benchmarking}. Machine learning pipelines are the key elements that let us evaluate these edge ML algorithms. Thus, we develop an end-to-end ML pipeline that facilities developing anomaly detection systems where the detection can be done on different layers (edge, fog, and cloud). We make the following observations based on our findings: (i) lack of complete multi-protocol edge anomaly detection frameworks, (ii) the lack of control over converted ML models, (iii) there is a lack of sensor data terminology, (iv) manufacturers/developers provide platform-specific support, (vi) there is lack of support for multi-language/protocol development frameworks, (vii) automated is actually not automated.

\textit{Lack of complete multi-protocol edge anomaly detection frameworks.} The two tasks that are executed within anomaly detection pipelines are: (i) data gathering/ingestion, (ii) anomaly detection model development. While these tasks are different by nature, they are essential to implement a complete IoT anomaly detection mechanism. During the evaluations, we see that the required tools are very diverse and no ML framework offers a complete solution. For data ingestion, flexibility and scalability are the most desired features. Hence, microcontrollers are the primary elements that are deployed in IoT environments to gather data. Due to the resource constraints of these environments, we need low power-consuming communication protocols. This is one of the main reasons that why BLE \cite{heydon2012bluetooth} is invented. In the edge, during the evaluations we see that there are three conditions needed for data acquiring: (i) sensors should be compatible with the microcontroller, (ii) communication protocol libraries should be available to establish a network both for the edge and the fog, and (iii) configurable platform that provides visual assistance is needed. As no framework ensures all three conditions, we had to use tools that require different programming languages (e.g., Arduino, JavaScript, Python).

\textit{The lack of control over converted ML models.} TensorFlow allows us to convert the neural network models that are generated via the main API to more lightweight models that can be deployed to edge and fog. However it offers zero control over the model conversion. During the evaluations, we realized that some lite models perform better than their main peers which were unexpected as there is a known trade-off between model accuracy and power consumption. Due to not having any control over this conversion, the only way to compare these models is by running inference. Also, scikit-learn models do not have a lite version. Pickle \cite{pickle} library is used to pack and deploy the same model to another platform. Hence deploying these models to microcontrollers is currently not possible. This is why during the evaluations we did not observe any differences rather than inference time when running scikit-learn models.

\begin{table}[!t]
\centering
\footnotesize
\caption{Data Format}
\label{tab:dataFormat}
\begin{adjustbox}{width = \textwidth}
\def\arraystretch{1.2}
\begin{threeparttable}
\begin{tabular}{@{}lcccccccccccc@{}}
\hline
Categories                        & \multicolumn{5}{c}{Sensor Type}                     & \multicolumn{4}{c}{Protocols}          & \multicolumn{3}{c}{Others}  \\ \hline
\multicolumn{1}{l|}{Features} &
  \multicolumn{1}{l}{Temperature} &
  \multicolumn{1}{l}{Humidity} &
  \multicolumn{1}{l}{Air Quality} &
  \multicolumn{1}{l}{Light} &
  \multicolumn{1}{l|}{Sound} &
  \multicolumn{1}{l}{WiFi} &
  \multicolumn{1}{l}{Bluetooth Classic} &
  \multicolumn{1}{l}{BLE} &
  \multicolumn{1}{l|}{Zigbee} &
  \multicolumn{1}{l}{Location ID} &
  \multicolumn{1}{l}{Sensor ID} &
  \multicolumn{1}{l}{Microcontroller ID} \\
\multicolumn{1}{l|}{Identifiers}  & TH    & HU    & AQ & LI  & \multicolumn{1}{c|}{SO}  & WF & BC & BL & \multicolumn{1}{c|}{ZB} & Numbers & Numbers & Numbers \\
\multicolumn{1}{l|}{Example Data} & 24.45 & 44.31 & 75 & 255 & \multicolumn{1}{c|}{644} & NA & NA & NA & \multicolumn{1}{c|}{NA} & 01      & 001     & 1       \\ \hline
\end{tabular}
\begin{tablenotes}
\setlength\labelsep{0pt}
 \small
 \item NA: Not Applicable. 
\end{tablenotes}
\end{threeparttable}
\end{adjustbox}
\end{table}

\textit{Lack of unified IoT sensor data terminology.} In an IoT environment, the sensors generate data in a similar format. Showing such similarities might be confusing if the data are not identified with certain terms/identifiers especially where big data are present. Hence, we apply further processing to identify the data context. However, there is no sensor terminology that explicitly states the sensor data format. Thus, we designed ECCG in a way that it generates a code that provides all required information such as the location of the sensor, sensor type, and the microcontroller model. Also, another important point to consider regarding IoT sensor data terminology is the data with floating points. While WiFi, Zigbee, and Bluetooth Classic have no issues when sending floating-point data, BLE by nature designed to send buffer (byte array) as it provides power optimization. Hence, when sending data over BLE we might require further indicators such as "F" for floats or "I" for integers to correctly identify the data. Table \ref{tab:dataFormat} presents the data format that is generated via ECCG. We set the number IDs in different lengths to prevent possible confusion.

We evaluated our AnoML pipeline with two datasets, WADI and AnoML. WADI dataset has data from 127 different sensors, which is not possible to be replayed on edge devices due to limited computing power. WADI dataset was only evaluated on fog and cloud. We used a private machine with NVidia Tesla GPU and TensorFlow-GPU library version 2.4.1 for ML model training. AnoML dataset has data from five environmental sensors but two of them (temperature and humidity) were used to build and evaluate models on edge, fog, and cloud. We used Google Colab with GPU support to build models for the AnoML dataset with TensorFlow-GPU library version 2.1.1. AnoML dataset was transformed using reduction techniques only, to convert it into univariate.

\begin{figure*}
\caption{WADI Dataset Evaluation}
\label{figall}
\begin{multicols}{2}
    \caption{Inference Time}
    \label{fig:WADI-InferenceTime}
    \includegraphics[width=\linewidth]{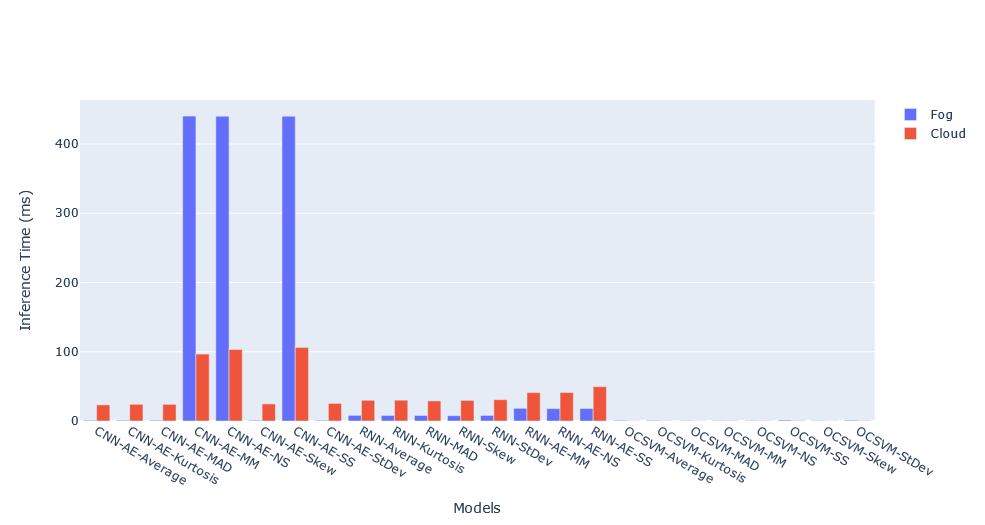}\par 
    \caption{Accuracy}
    \label{fig:WADI-Accuracy}
    \includegraphics[width=\linewidth]{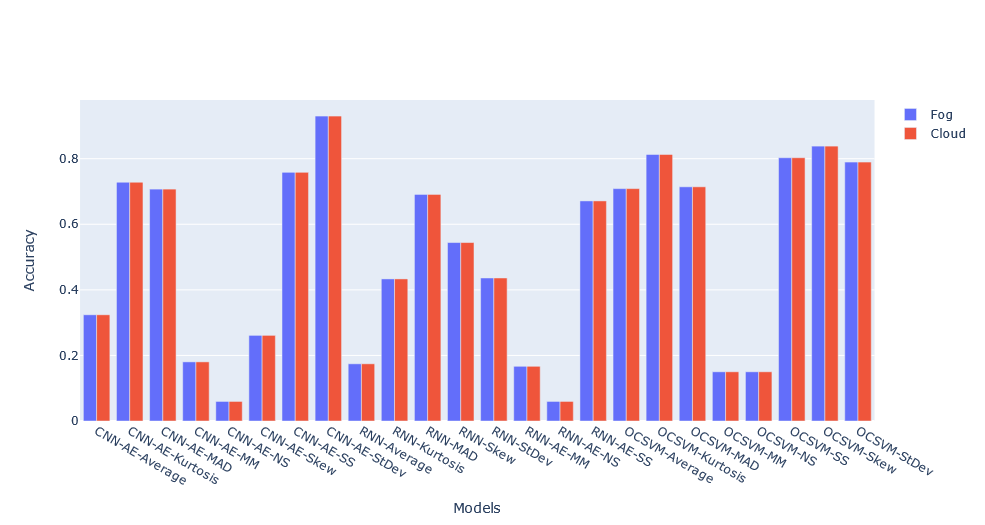}\par
    \end{multicols}
\begin{multicols}{2}
    \caption{AUC}
    \label{fig:WADI-AUC}
    \includegraphics[width=\linewidth]{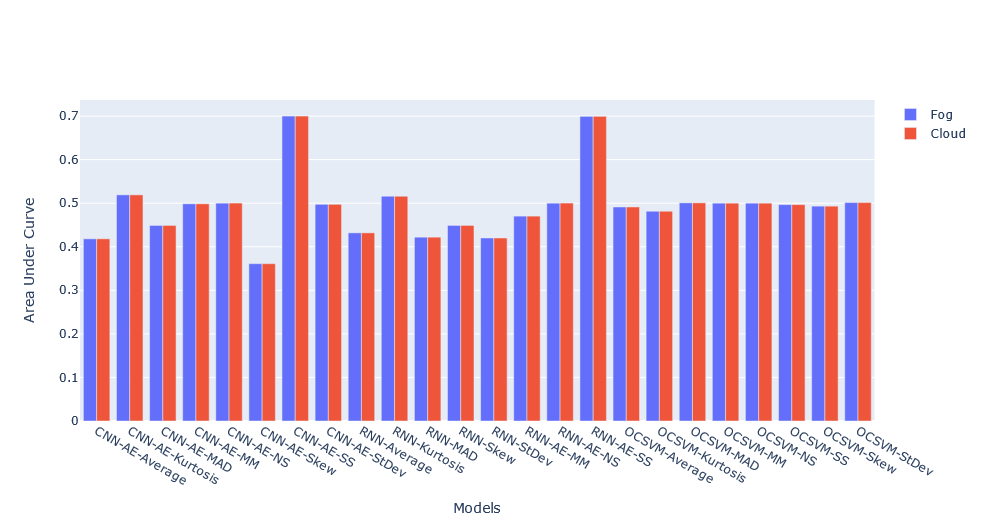}\par
    \caption{F1-Score}
    \label{fig:WADI-F1Score}
    \includegraphics[width=\linewidth]{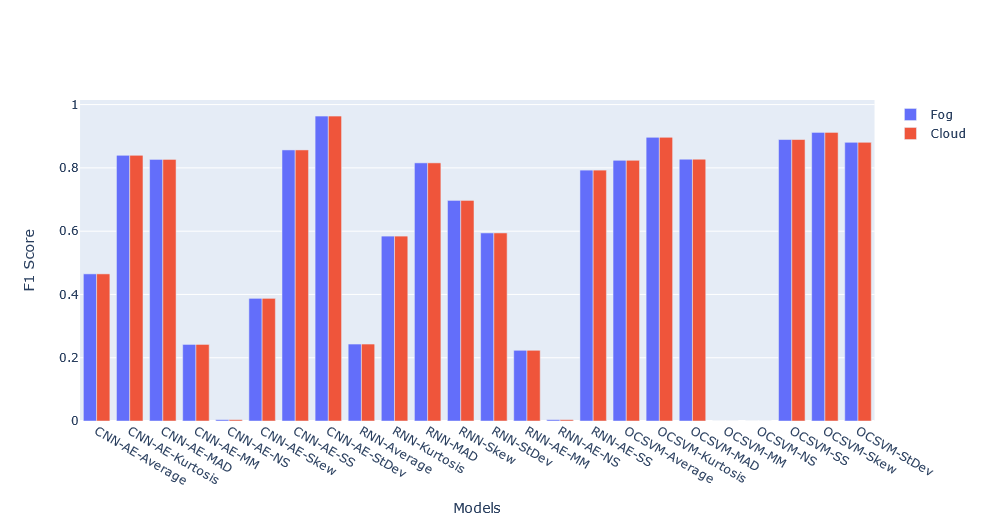}\par
\end{multicols}
\begin{multicols}{2}
    \caption{Precision}
    \label{fig:WADI-Precision}
    \includegraphics[width=\linewidth]{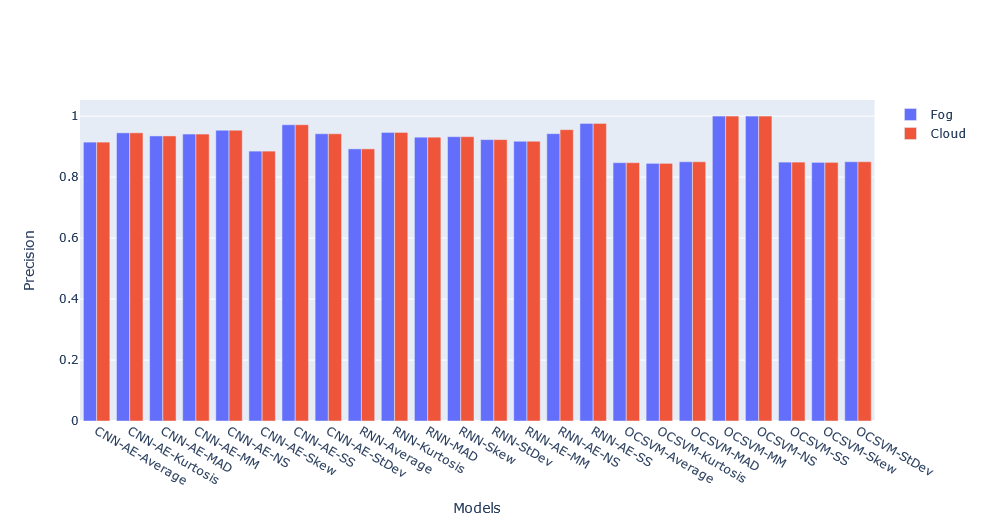}\par
    \caption{Recall}
    \label{fig:WADI-Recal}
    \includegraphics[width=\linewidth]{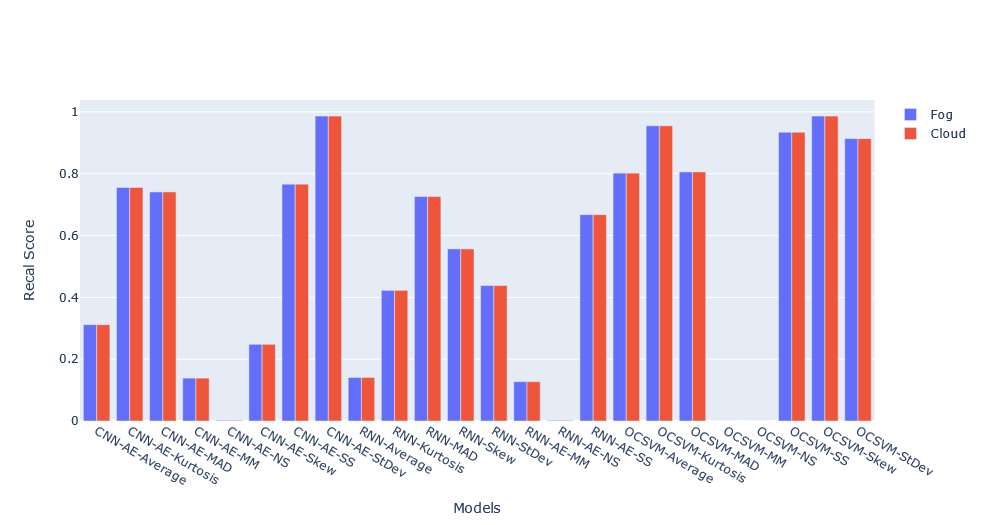}\par
    \end{multicols}
\begin{multicols}{2}
    \caption{Scaling Time}
    \label{fig:WADI-SRTime}
    \includegraphics[width=\linewidth]{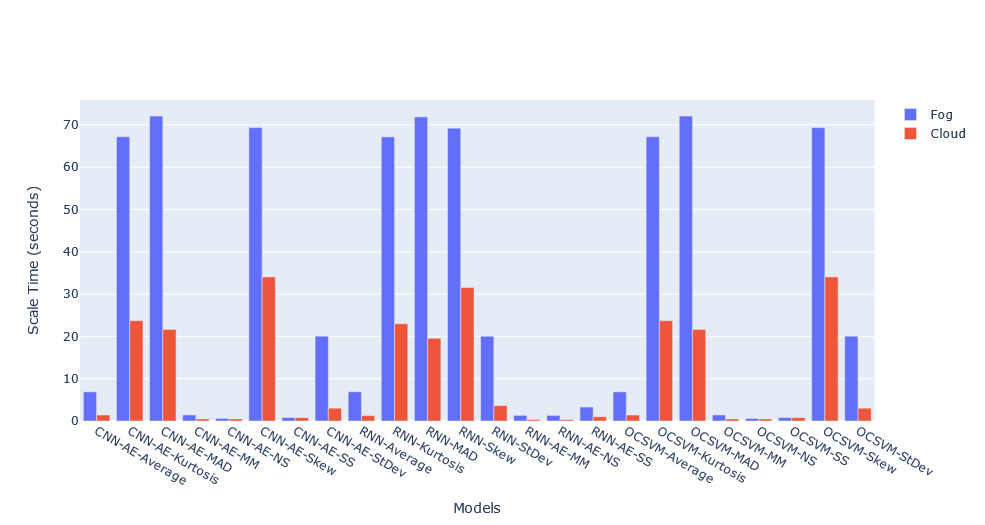}\par
    \caption{Model Size}
    \label{fig:WADI-sizeondisk}
    \includegraphics[width=\linewidth]{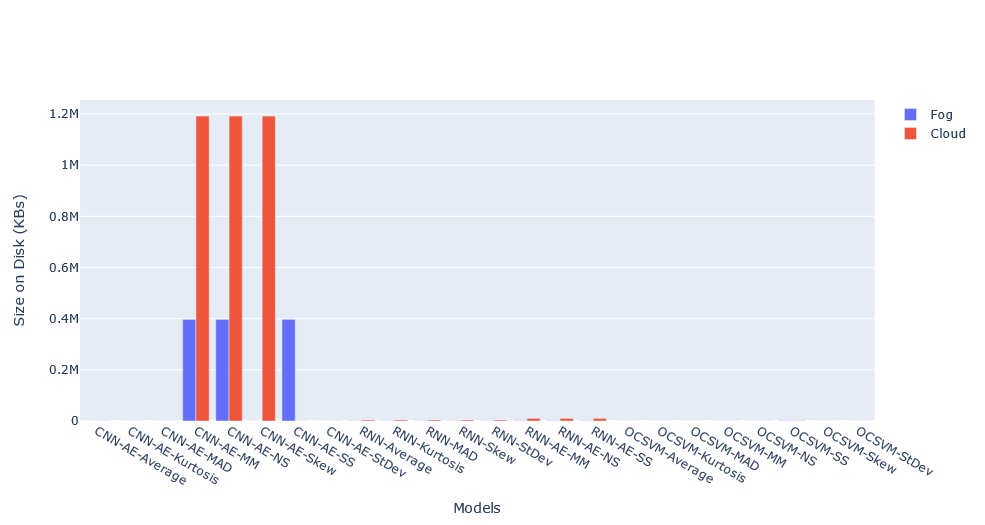}\par
\end{multicols}
\end{figure*}

In terms of accuracy, F1-score, precision, recall, and Area Under Curve (AUC), there was no significant difference recorded when comparing fog and cloud inferences in all types of models for the WADI Dataset. It was also observed that batch processing (all at once) in TensorFlow on the cloud does not affect the efficiency of the pipeline. Figure \ref{figall} shows the comparison of each metric for all models for the WADI Dataset. Regarding fog and cloud, we observed a similar trend for the AnoML Dataset. However, we noticed that edge results were comparatively less efficient.

We noticed that inference on fog takes distinctly less time as compared to inference on the cloud. But, we observed the opposite results for CNN-AE when doing multivariate analysis where we recorded that cloud inference was faster than fog. Figure \ref{fig:WADI-InferenceTime} provides a visual comparison between inference times of fog and cloud. We also experienced that when using batch prediction in the cloud (all rows at once), the average prediction time was extremely faster. But, this is not applicable in the real-world scenario where data is being streamed. Both scaling and reduction techniques were faster on cloud (using batch-prediction) when compared to fog as seen in Figure \ref{fig:WADI-SRTime}. Inference time trends were also identical in ML models for the AnoML dataset as fog inference time was less than cloud inference. The fog was more than 10 times faster than cloud inference. Inference time on edge was extremely slow when compared to fog and cloud, it was more than 100 times slower than the fog and almost 5 times slower than the cloud.

WADI Dataset is based on two sub-datasets recorded at different times, normal and attack, as we discussed that normal dataset was used for ML model training thus we used the attack dataset for testing. The attack dataset consists of 172800 rows which were then converted into time-series data. As a result, the dimension of the data became (172770, 30, 127) for scale-based and (172770, 30, 1) for reduction-based models. Reduction-based models took exceptionally less time for training because of the univariate nature of the data. The size-on-disk of CNN-AE scaled-based models is too big as compared to CNN-AE reduction-based models. For RNN-AE models, the size-on-disk of reduction-based models are not significantly different from scale-based RNN models. The main reason for this was that we used LSTM layers for RNN. The size-on-disk variation trend was identical in fog models as seen in Figure \ref{fig:WADI-sizeondisk}. We also learned that scikit-learn models maintain consistency over fog and cloud. The time-related results in the fog were always slower due to the difference in computing power, but there was no change observed in accuracy-related metrics. The obvious reason was that there was no platform/format conversion done for scikit-learn models.

Models for the AnoML dataset were lightweight but there was a notable difference for the fog models in cloud model size-on-disk. We also observed that the size of edge models was significantly greater than fog models, even though these models were converted from them. The reason was that edge models were in plain-text (hex-dump) as they were a C-array, but fog models were flat-buffered-binary format.

\section{Conclusion}
\label{Sec:Conclusion}
The proposed system offers support to a data scientist with minimal IoT knowledge to deploy a reconfigurable IoT anomaly detection infrastructure that utilizes a data science pipeline. The proposed framework contains: edge nodes consist of microcontrollers, fog nodes consist of single board computers and virtual cloud nodes. The communication between nodes is not limited to edge node types but limited with protocol specifications (e.g., Bluetooth only allows seven slaves/servers), hence allowing the implementation of different network topologies (e.g., bus, tree, and star). The system also explicitly supports the four major phases of data processing: gathering, training, deployment, and inference. We evaluated several combinations to measure the scalability that is offered by the proposed framework. We  observed that the proposed anomaly detection pipeline reduces the required labor for building an IoT anomaly detection system. We identified the drawbacks of the proposed system including the reasons behind them. We believe our work encourages the future studies that aim to build open-source ML-based pipelines. 

As future work, we are seeking to improve the ECCG web application by including more edge nodes, sensors, and protocol types. In addition, we envision converting required manual processes (e.g., deploying ML model to edge, uploading training data, deploying edge code) within the AnoML-IoT to automated.

\bibliographystyle{elsarticle-num-names}
\bibliography{sample}

\begin{thebibliography}{82}
\expandafter\ifx\csname natexlab\endcsname\relax\def\natexlab#1{#1}\fi
\providecommand{\url}[1]{\texttt{#1}}
\providecommand{\href}[2]{#2}
\providecommand{\path}[1]{#1}
\providecommand{\DOIprefix}{doi:}
\providecommand{\ArXivprefix}{arXiv:}
\providecommand{\URLprefix}{URL: }
\providecommand{\Pubmedprefix}{pmid:}
\providecommand{\doi}[1]{\href{http://dx.doi.org/#1}{\path{#1}}}
\providecommand{\Pubmed}[1]{\href{pmid:#1}{\path{#1}}}
\providecommand{\bibinfo}[2]{#2}
\ifx\xfnm\relax \def\xfnm[#1]{\unskip,\space#1}\fi
\bibitem[{Lasi et~al.(2014)Lasi, Fettke, Kemper, Feld, and
  Hoffmann}]{lasi2014industry}
\bibinfo{author}{H.~Lasi}, \bibinfo{author}{P.~Fettke}, \bibinfo{author}{H.-G.
  Kemper}, \bibinfo{author}{T.~Feld}, \bibinfo{author}{M.~Hoffmann},
\newblock \bibinfo{title}{Industry 4.0},
\newblock \bibinfo{journal}{Business \& information systems engineering}
  \bibinfo{volume}{6} (\bibinfo{year}{2014}) \bibinfo{pages}{239--242}.
\bibitem[{Mohammadi et~al.(2018)Mohammadi, Al-Fuqaha, Sorour, and
  Guizani}]{Mohammadi2018}
\bibinfo{author}{M.~Mohammadi}, \bibinfo{author}{A.~Al-Fuqaha},
  \bibinfo{author}{S.~Sorour}, \bibinfo{author}{M.~Guizani},
\newblock \bibinfo{title}{{Deep learning for IoT big data and streaming
  analytics: A survey}},
\newblock \bibinfo{journal}{IEEE Commun. Surv. Tutorials} \bibinfo{volume}{20}
  (\bibinfo{year}{2018}) \bibinfo{pages}{2923--2960}.
  \DOIprefix\doi{10.1109/COMST.2018.2844341}.
  \href{http://arxiv.org/abs/1712.04301}{{\tt arXiv:1712.04301}}.
\bibitem[{Jiang and Li(2020)}]{jiang2020convolutional}
\bibinfo{author}{Y.~Jiang}, \bibinfo{author}{C.~Li},
\newblock \bibinfo{title}{Convolutional neural networks for image-based
  high-throughput plant phenotyping: a review},
\newblock \bibinfo{journal}{Plant Phenomics} \bibinfo{volume}{2020}
  (\bibinfo{year}{2020}).
\bibitem[{Goh et~al.(2017)Goh, Adepu, Tan, and Lee}]{goh2017anomaly}
\bibinfo{author}{J.~Goh}, \bibinfo{author}{S.~Adepu}, \bibinfo{author}{M.~Tan},
  \bibinfo{author}{Z.~S. Lee},
\newblock \bibinfo{title}{Anomaly detection in cyber physical systems using
  recurrent neural networks},
\newblock in: \bibinfo{booktitle}{2017 IEEE 18th International Symposium on
  High Assurance Systems Engineering (HASE)}, \bibinfo{organization}{IEEE},
  \bibinfo{year}{2017}, pp. \bibinfo{pages}{140--145}.
\bibitem[{Liu et~al.(2008)Liu, Ting, and Zhou}]{liu2008isolation}
\bibinfo{author}{F.~T. Liu}, \bibinfo{author}{K.~M. Ting},
  \bibinfo{author}{Z.-H. Zhou},
\newblock \bibinfo{title}{Isolation forest},
\newblock in: \bibinfo{booktitle}{2008 eighth ieee international conference on
  data mining}, \bibinfo{organization}{IEEE}, \bibinfo{year}{2008}, pp.
  \bibinfo{pages}{413--422}.
\bibitem[{Ma and Perkins(2003)}]{ma2003time}
\bibinfo{author}{J.~Ma}, \bibinfo{author}{S.~Perkins},
\newblock \bibinfo{title}{Time-series novelty detection using one-class support
  vector machines},
\newblock in: \bibinfo{booktitle}{Proceedings of the International Joint
  Conference on Neural Networks, 2003.}, volume~\bibinfo{volume}{3},
  \bibinfo{organization}{IEEE}, \bibinfo{year}{2003}, pp.
  \bibinfo{pages}{1741--1745}.
\bibitem[{Kayan(2021)}]{dataset}
\bibinfo{author}{H.~Kayan}, \bibinfo{title}{{AnoML}-{IoT}},
  \bibinfo{year}{2021}. \URLprefix \url{https://kaggle.com/hkayan/anomliot}.
\bibitem[{Ahmed et~al.(2017)Ahmed, Palleti, and Mathur}]{ahmed2017wadi}
\bibinfo{author}{C.~M. Ahmed}, \bibinfo{author}{V.~R. Palleti},
  \bibinfo{author}{A.~P. Mathur},
\newblock \bibinfo{title}{Wadi: a water distribution testbed for research in
  the design of secure cyber physical systems},
\newblock in: \bibinfo{booktitle}{Proceedings of the 3rd International Workshop
  on Cyber-Physical Systems for Smart Water Networks}, \bibinfo{year}{2017},
  pp. \bibinfo{pages}{25--28}.
\bibitem[{Liu et~al.(2020)Liu, Pang, Karlsson, and Gong}]{Liu2020}
\bibinfo{author}{Y.~Liu}, \bibinfo{author}{Z.~Pang},
  \bibinfo{author}{M.~Karlsson}, \bibinfo{author}{S.~Gong},
\newblock \bibinfo{title}{{Anomaly detection based on machine learning in
  IoT-based vertical plant wall for indoor climate control}},
\newblock \bibinfo{journal}{Build. Environ.} \bibinfo{volume}{183}
  (\bibinfo{year}{2020}). \DOIprefix\doi{10.1016/j.buildenv.2020.107212}.
\bibitem[{Chalapathy and Chawla(2019)}]{Chalapathy2019}
\bibinfo{author}{R.~Chalapathy}, \bibinfo{author}{S.~Chawla},
\newblock \bibinfo{title}{{Deep learning for anomaly detection: A survey}},
\newblock \bibinfo{journal}{arXiv}  (\bibinfo{year}{2019})
  \bibinfo{pages}{1--50}. \href{http://arxiv.org/abs/1901.03407}{{\tt
  arXiv:1901.03407}}.
\bibitem[{Bl{\'{a}}zquez-Garc{\'{i}}a et~al.(2020)Bl{\'{a}}zquez-Garc{\'{i}}a,
  Conde, Mori, and Lozano}]{Blazquez-Garcia2020}
\bibinfo{author}{A.~Bl{\'{a}}zquez-Garc{\'{i}}a}, \bibinfo{author}{A.~Conde},
  \bibinfo{author}{U.~Mori}, \bibinfo{author}{J.~A. Lozano},
\newblock \bibinfo{title}{{A review on outlier/anomaly detection in time series
  data}},
\newblock \bibinfo{journal}{arXiv}  (\bibinfo{year}{2020}).
  \href{http://arxiv.org/abs/2002.04236}{{\tt arXiv:2002.04236}}.
\bibitem[{Song et~al.(2013)Song, Takakura, Okabe, and Nakao}]{song2013toward}
\bibinfo{author}{J.~Song}, \bibinfo{author}{H.~Takakura},
  \bibinfo{author}{Y.~Okabe}, \bibinfo{author}{K.~Nakao},
\newblock \bibinfo{title}{Toward a more practical unsupervised anomaly
  detection system},
\newblock \bibinfo{journal}{Information Sciences} \bibinfo{volume}{231}
  (\bibinfo{year}{2013}) \bibinfo{pages}{4--14}.
\bibitem[{Zhu et~al.(2012)Zhu, Nayak, and Roy-Chowdhury}]{zhu2012context}
\bibinfo{author}{Y.~Zhu}, \bibinfo{author}{N.~M. Nayak}, \bibinfo{author}{A.~K.
  Roy-Chowdhury},
\newblock \bibinfo{title}{Context-aware activity recognition and anomaly
  detection in video},
\newblock \bibinfo{journal}{IEEE Journal of Selected Topics in Signal
  Processing} \bibinfo{volume}{7} (\bibinfo{year}{2012})
  \bibinfo{pages}{91--101}.
\bibitem[{Mart{\'\i} et~al.(2015)Mart{\'\i}, Sanchez-Pi, Molina, and
  Garcia}]{marti2015anomaly}
\bibinfo{author}{L.~Mart{\'\i}}, \bibinfo{author}{N.~Sanchez-Pi},
  \bibinfo{author}{J.~M. Molina}, \bibinfo{author}{A.~C.~B. Garcia},
\newblock \bibinfo{title}{Anomaly detection based on sensor data in petroleum
  industry applications},
\newblock \bibinfo{journal}{Sensors} \bibinfo{volume}{15}
  (\bibinfo{year}{2015}) \bibinfo{pages}{2774--2797}.
\bibitem[{Jazdi(2014)}]{jazdi2014cyber}
\bibinfo{author}{N.~Jazdi},
\newblock \bibinfo{title}{Cyber physical systems in the context of industry
  4.0},
\newblock in: \bibinfo{booktitle}{2014 IEEE international conference on
  automation, quality and testing, robotics}, \bibinfo{organization}{IEEE},
  \bibinfo{year}{2014}, pp. \bibinfo{pages}{1--4}.
\bibitem[{Raghavendra et~al.(2006)Raghavendra, Sivalingam, and
  Znati}]{raghavendra2006wireless}
\bibinfo{author}{C.~S. Raghavendra}, \bibinfo{author}{K.~M. Sivalingam},
  \bibinfo{author}{T.~Znati}, \bibinfo{title}{Wireless sensor networks},
  \bibinfo{publisher}{Springer}, \bibinfo{year}{2006}.
\bibitem[{Pang et~al.(2017)Pang, Liu, Peng, and Peng}]{pang2017anomaly}
\bibinfo{author}{J.~Pang}, \bibinfo{author}{D.~Liu}, \bibinfo{author}{Y.~Peng},
  \bibinfo{author}{X.~Peng},
\newblock \bibinfo{title}{Anomaly detection based on uncertainty fusion for
  univariate monitoring series},
\newblock \bibinfo{journal}{Measurement} \bibinfo{volume}{95}
  (\bibinfo{year}{2017}) \bibinfo{pages}{280--292}.
\bibitem[{Su et~al.(2019)Su, Zhao, Niu, Liu, Sun, and Pei}]{su2019robust}
\bibinfo{author}{Y.~Su}, \bibinfo{author}{Y.~Zhao}, \bibinfo{author}{C.~Niu},
  \bibinfo{author}{R.~Liu}, \bibinfo{author}{W.~Sun}, \bibinfo{author}{D.~Pei},
\newblock \bibinfo{title}{Robust anomaly detection for multivariate time series
  through stochastic recurrent neural network},
\newblock in: \bibinfo{booktitle}{Proceedings of the 25th ACM SIGKDD
  International Conference on Knowledge Discovery \& Data Mining},
  \bibinfo{year}{2019}, pp. \bibinfo{pages}{2828--2837}.
\bibitem[{Teng(2010)}]{Teng2010}
\bibinfo{author}{M.~Teng},
\newblock \bibinfo{title}{{Anomaly detection on time series}},
\newblock \bibinfo{journal}{Proc. 2010 IEEE Int. Conf. Prog. Informatics
  Comput. PIC 2010} \bibinfo{volume}{1} (\bibinfo{year}{2010})
  \bibinfo{pages}{603--608}. \DOIprefix\doi{10.1109/PIC.2010.5687485}.
\bibitem[{Wu(2017)}]{Wu2017}
\bibinfo{author}{H.~S. Wu},
\newblock \bibinfo{title}{{A survey of research on anomaly detection for time
  series}},
\newblock \bibinfo{journal}{2016 13th Int. Comput. Conf. Wavelet Act. Media
  Technol. Inf. Process. ICCWAMTIP 2017}  (\bibinfo{year}{2017})
  \bibinfo{pages}{426--431}. \DOIprefix\doi{10.1109/ICCWAMTIP.2016.8079887}.
\bibitem[{Chandola et~al.(2009)Chandola, Banerjee, and
  Kumar}]{chandola2009anomaly}
\bibinfo{author}{V.~Chandola}, \bibinfo{author}{A.~Banerjee},
  \bibinfo{author}{V.~Kumar},
\newblock \bibinfo{title}{Anomaly detection: A survey},
\newblock \bibinfo{journal}{ACM computing surveys (CSUR)} \bibinfo{volume}{41}
  (\bibinfo{year}{2009}) \bibinfo{pages}{1--58}.
\bibitem[{Pedregosa et~al.(2011)Pedregosa, Varoquaux, Gramfort, Michel,
  Thirion, Grisel, Blondel, Prettenhofer, Weiss, Dubourg
  et~al.}]{pedregosa2011scikit}
\bibinfo{author}{F.~Pedregosa}, \bibinfo{author}{G.~Varoquaux},
  \bibinfo{author}{A.~Gramfort}, \bibinfo{author}{V.~Michel},
  \bibinfo{author}{B.~Thirion}, \bibinfo{author}{O.~Grisel},
  \bibinfo{author}{M.~Blondel}, \bibinfo{author}{P.~Prettenhofer},
  \bibinfo{author}{R.~Weiss}, \bibinfo{author}{V.~Dubourg}, et~al.,
\newblock \bibinfo{title}{Scikit-learn: Machine learning in python},
\newblock \bibinfo{journal}{the Journal of machine Learning research}
  \bibinfo{volume}{12} (\bibinfo{year}{2011}) \bibinfo{pages}{2825--2830}.
\bibitem[{Abadi(2016)}]{abadi2016tensorflow}
\bibinfo{author}{M.~Abadi},
\newblock \bibinfo{title}{Tensorflow: learning functions at scale},
\newblock in: \bibinfo{booktitle}{Proceedings of the 21st ACM SIGPLAN
  International Conference on Functional Programming}, \bibinfo{year}{2016},
  pp. \bibinfo{pages}{1--1}.
\bibitem[{Munir et~al.(2018)Munir, Siddiqui, Dengel, and
  Ahmed}]{munir2018deepant}
\bibinfo{author}{M.~Munir}, \bibinfo{author}{S.~A. Siddiqui},
  \bibinfo{author}{A.~Dengel}, \bibinfo{author}{S.~Ahmed},
\newblock \bibinfo{title}{Deepant: A deep learning approach for unsupervised
  anomaly detection in time series},
\newblock \bibinfo{journal}{IEEE Access} \bibinfo{volume}{7}
  (\bibinfo{year}{2018}) \bibinfo{pages}{1991--2005}.
\bibitem[{Wen and Keyes(2019)}]{wen2019time}
\bibinfo{author}{T.~Wen}, \bibinfo{author}{R.~Keyes},
\newblock \bibinfo{title}{Time series anomaly detection using convolutional
  neural networks and transfer learning},
\newblock \bibinfo{journal}{arXiv preprint arXiv:1905.13628}
  (\bibinfo{year}{2019}).
\bibitem[{Li et~al.(2019)Li, Chen, Jin, Shi, Goh, and Ng}]{li2019mad}
\bibinfo{author}{D.~Li}, \bibinfo{author}{D.~Chen}, \bibinfo{author}{B.~Jin},
  \bibinfo{author}{L.~Shi}, \bibinfo{author}{J.~Goh}, \bibinfo{author}{S.-K.
  Ng},
\newblock \bibinfo{title}{Mad-gan: Multivariate anomaly detection for time
  series data with generative adversarial networks},
\newblock in: \bibinfo{booktitle}{International Conference on Artificial Neural
  Networks}, \bibinfo{organization}{Springer}, \bibinfo{year}{2019}, pp.
  \bibinfo{pages}{703--716}.
\bibitem[{TensorFlow(2021)}]{TF_decision}
\bibinfo{author}{TensorFlow}, \bibinfo{title}{Introducing {TensorFlow}
  {Decision} {Forests}}, \bibinfo{year}{2021}. \URLprefix
  \url{https://blog.tensorflow.org/}.
\bibitem[{Ketkar(2017)}]{ketkar2017introduction}
\bibinfo{author}{N.~Ketkar},
\newblock \bibinfo{title}{Introduction to pytorch},
\newblock in: \bibinfo{booktitle}{Deep learning with python},
  \bibinfo{publisher}{Springer}, \bibinfo{year}{2017}, pp.
  \bibinfo{pages}{195--208}.
\bibitem[{QuinnRadich(2021)}]{quinnradich_automatic}
\bibinfo{author}{QuinnRadich}, \bibinfo{title}{Automatic code generation with
  mlgen}, \bibinfo{year}{2021}. \URLprefix \url{https://docs.microsoft.com/}.
\bibitem[{MATLAB(2021)}]{MATLABML}
\bibinfo{author}{MATLAB}, \bibinfo{title}{Deep {Learning} {Code} {Generation} -
  {MATLAB} \& {Simulink} - {MathWorks} {United} {Kingdom}},
  \bibinfo{year}{2021}. \URLprefix
  \url{https://uk.mathworks.com/help/deeplearning/deep-learning-code-generation.html}.
\bibitem[{Ren et~al.(2019)Ren, Xu, Wang, Yi, Huang, Kou, Xing, Yang, Tong, and
  Zhang}]{ren2019time}
\bibinfo{author}{H.~Ren}, \bibinfo{author}{B.~Xu}, \bibinfo{author}{Y.~Wang},
  \bibinfo{author}{C.~Yi}, \bibinfo{author}{C.~Huang},
  \bibinfo{author}{X.~Kou}, \bibinfo{author}{T.~Xing},
  \bibinfo{author}{M.~Yang}, \bibinfo{author}{J.~Tong},
  \bibinfo{author}{Q.~Zhang},
\newblock \bibinfo{title}{Time-series anomaly detection service at microsoft},
\newblock in: \bibinfo{booktitle}{Proceedings of the 25th ACM SIGKDD
  International Conference on Knowledge Discovery \& Data Mining},
  \bibinfo{year}{2019}, pp. \bibinfo{pages}{3009--3017}.
\bibitem[{NilsPohlmann(2021)}]{nilspohlmann_create}
\bibinfo{author}{NilsPohlmann}, \bibinfo{title}{Create and run {ML} pipelines -
  {Azure} {Machine} {Learning}}, \bibinfo{year}{2021}. \URLprefix
  \url{https://docs.microsoft.com/en-us/azure/machine-learning/}.
\bibitem[{Microsoft(2021{\natexlab{a}})}]{cognitive}
\bibinfo{author}{Microsoft}, \bibinfo{title}{Cognitive {Services} – {APIs}
  for {AI} {Developers} {\textbar} {Microsoft} {Azure}},
  \bibinfo{year}{2021}{\natexlab{a}}. \URLprefix
  \url{https://azure.microsoft.com/en-gb/services/cognitive-services/}.
\bibitem[{Microsoft(2021{\natexlab{b}})}]{azureAnomaly}
\bibinfo{author}{Microsoft}, \bibinfo{title}{Anomaly {Detector} - {Anomaly}
  {Detection} {System} {\textbar} {Microsoft} {Azure}},
  \bibinfo{year}{2021}{\natexlab{b}}. \URLprefix
  \url{https://azure.microsoft.com/en-us/services/cognitive-services/anomaly-detector/}.
\bibitem[{Zhao et~al.(2020)Zhao, Wang, Duan, Huang, Cao, Tong, Xu, Bai, Tong,
  and Zhang}]{zhao2020multivariate}
\bibinfo{author}{H.~Zhao}, \bibinfo{author}{Y.~Wang},
  \bibinfo{author}{J.~Duan}, \bibinfo{author}{C.~Huang},
  \bibinfo{author}{D.~Cao}, \bibinfo{author}{Y.~Tong}, \bibinfo{author}{B.~Xu},
  \bibinfo{author}{J.~Bai}, \bibinfo{author}{J.~Tong},
  \bibinfo{author}{Q.~Zhang},
\newblock \bibinfo{title}{Multivariate time-series anomaly detection via graph
  attention network},
\newblock \bibinfo{journal}{arXiv preprint arXiv:2009.02040}
  (\bibinfo{year}{2020}).
\bibitem[{AWS(2021)}]{aws_build_2021}
\bibinfo{author}{AWS},
\newblock \bibinfo{title}{Build your own {Anomaly} {Detection} {ML} {Pipeline}}
   (\bibinfo{year}{2021}) \bibinfo{pages}{1}.
\bibitem[{Liberty et~al.(2020)Liberty, Karnin, Xiang, Rouesnel, Coskun,
  Nallapati, Delgado, Sadoughi, Astashonok, Das et~al.}]{liberty2020elastic}
\bibinfo{author}{E.~Liberty}, \bibinfo{author}{Z.~Karnin},
  \bibinfo{author}{B.~Xiang}, \bibinfo{author}{L.~Rouesnel},
  \bibinfo{author}{B.~Coskun}, \bibinfo{author}{R.~Nallapati},
  \bibinfo{author}{J.~Delgado}, \bibinfo{author}{A.~Sadoughi},
  \bibinfo{author}{Y.~Astashonok}, \bibinfo{author}{P.~Das}, et~al.,
\newblock \bibinfo{title}{Elastic machine learning algorithms in amazon
  sagemaker},
\newblock in: \bibinfo{booktitle}{Proceedings of the 2020 ACM SIGMOD
  International Conference on Management of Data}, \bibinfo{year}{2020}, pp.
  \bibinfo{pages}{731--737}.
\bibitem[{Guha et~al.(2016)Guha, Mishra, Roy, and Schrijvers}]{guha2016robust}
\bibinfo{author}{S.~Guha}, \bibinfo{author}{N.~Mishra},
  \bibinfo{author}{G.~Roy}, \bibinfo{author}{O.~Schrijvers},
\newblock \bibinfo{title}{Robust random cut forest based anomaly detection on
  streams},
\newblock in: \bibinfo{booktitle}{International conference on machine
  learning}, \bibinfo{organization}{PMLR}, \bibinfo{year}{2016}, pp.
  \bibinfo{pages}{2712--2721}.
\bibitem[{Prado et~al.(2020)Prado, Su, Saeed, Keller, Vallez, Anderson, Gregg,
  Benini, Llewellynn, Ouerhani et~al.}]{prado2020bonseyes}
\bibinfo{author}{M.~D. Prado}, \bibinfo{author}{J.~Su},
  \bibinfo{author}{R.~Saeed}, \bibinfo{author}{L.~Keller},
  \bibinfo{author}{N.~Vallez}, \bibinfo{author}{A.~Anderson},
  \bibinfo{author}{D.~Gregg}, \bibinfo{author}{L.~Benini},
  \bibinfo{author}{T.~Llewellynn}, \bibinfo{author}{N.~Ouerhani}, et~al.,
\newblock \bibinfo{title}{Bonseyes ai pipeline—bringing ai to you: End-to-end
  integration of data, algorithms, and deployment tools},
\newblock \bibinfo{journal}{ACM Transactions on Internet of Things}
  \bibinfo{volume}{1} (\bibinfo{year}{2020}) \bibinfo{pages}{1--25}.
\bibitem[{BAIR(2021)}]{caffe}
\bibinfo{author}{BAIR}, \bibinfo{title}{Caffe {\textbar} {Deep} {Learning}
  {Framework}}, \bibinfo{year}{2021}. \URLprefix
  \url{http://caffe.berkeleyvision.org/}.
\bibitem[{Foundation(WARE)}]{fiware}
\bibinfo{author}{T.~F. Foundation}, \bibinfo{title}{The {Open} {Source}
  platform for our smart digital future}, \bibinfo{year}{FIWARE}. \URLprefix
  \url{https://www.fiware.org/}.
\bibitem[{Fern{\'a}ndez et~al.(2013)Fern{\'a}ndez, D{\'\i}az, Mej{\'\i}as,
  L{\'o}pez, and Santos}]{fernandez2013kurento}
\bibinfo{author}{L.~L. Fern{\'a}ndez}, \bibinfo{author}{M.~P. D{\'\i}az},
  \bibinfo{author}{R.~B. Mej{\'\i}as}, \bibinfo{author}{F.~J. L{\'o}pez},
  \bibinfo{author}{J.~A. Santos},
\newblock \bibinfo{title}{Kurento: a media server technology for convergent
  www/mobile real-time multimedia communications supporting webrtc},
\newblock in: \bibinfo{booktitle}{2013 IEEE 14th International Symposium on" A
  World of Wireless, Mobile and Multimedia Networks"(WoWMoM)},
  \bibinfo{organization}{IEEE}, \bibinfo{year}{2013}, pp.
  \bibinfo{pages}{1--6}.
\bibitem[{TensorFlow(2021)}]{tensorflowLite}
\bibinfo{author}{TensorFlow}, \bibinfo{title}{{TensorFlow} {Lite} {\textbar}
  {ML} for {Mobile} and {Edge} {Devices}}, \bibinfo{year}{2021}. \URLprefix
  \url{https://www.tensorflow.org/lite}.
\bibitem[{Drori et~al.(2018)Drori, Krishnamurthy, Rampin, Louren{\c{c}}o, One,
  Cho, Silva, and Freire}]{drori2018alphad3m}
\bibinfo{author}{I.~Drori}, \bibinfo{author}{Y.~Krishnamurthy},
  \bibinfo{author}{R.~Rampin}, \bibinfo{author}{R.~Louren{\c{c}}o},
  \bibinfo{author}{J.~One}, \bibinfo{author}{K.~Cho},
  \bibinfo{author}{C.~Silva}, \bibinfo{author}{J.~Freire},
\newblock \bibinfo{title}{Alphad3m: Machine learning pipeline synthesis},
\newblock in: \bibinfo{booktitle}{AutoML Workshop at ICML},
  \bibinfo{year}{2018}.
\bibitem[{Browne et~al.(2012)Browne, Powley, Whitehouse, Lucas, Cowling,
  Rohlfshagen, Tavener, Perez, Samothrakis, and Colton}]{browne2012survey}
\bibinfo{author}{C.~B. Browne}, \bibinfo{author}{E.~Powley},
  \bibinfo{author}{D.~Whitehouse}, \bibinfo{author}{S.~M. Lucas},
  \bibinfo{author}{P.~I. Cowling}, \bibinfo{author}{P.~Rohlfshagen},
  \bibinfo{author}{S.~Tavener}, \bibinfo{author}{D.~Perez},
  \bibinfo{author}{S.~Samothrakis}, \bibinfo{author}{S.~Colton},
\newblock \bibinfo{title}{A survey of monte carlo tree search methods},
\newblock \bibinfo{journal}{IEEE Transactions on Computational Intelligence and
  AI in games} \bibinfo{volume}{4} (\bibinfo{year}{2012})
  \bibinfo{pages}{1--43}.
\bibitem[{Bottou(2010)}]{bottou2010large}
\bibinfo{author}{L.~Bottou},
\newblock \bibinfo{title}{Large-scale machine learning with stochastic gradient
  descent},
\newblock in: \bibinfo{booktitle}{Proceedings of COMPSTAT'2010},
  \bibinfo{publisher}{Springer}, \bibinfo{year}{2010}, pp.
  \bibinfo{pages}{177--186}.
\bibitem[{Sutton et~al.(2018)Sutton, Mahajan, Akbilgic, and
  Kamaleswaran}]{sutton2018physonline}
\bibinfo{author}{J.~R. Sutton}, \bibinfo{author}{R.~Mahajan},
  \bibinfo{author}{O.~Akbilgic}, \bibinfo{author}{R.~Kamaleswaran},
\newblock \bibinfo{title}{Physonline: an open source machine learning pipeline
  for real-time analysis of streaming physiological waveform},
\newblock \bibinfo{journal}{IEEE journal of biomedical and health informatics}
  \bibinfo{volume}{23} (\bibinfo{year}{2018}) \bibinfo{pages}{59--65}.
\bibitem[{Meng et~al.(2016)Meng, Bradley, Yavuz, Sparks, Venkataraman, Liu,
  Freeman, Tsai, Amde, Owen et~al.}]{meng2016mllib}
\bibinfo{author}{X.~Meng}, \bibinfo{author}{J.~Bradley},
  \bibinfo{author}{B.~Yavuz}, \bibinfo{author}{E.~Sparks},
  \bibinfo{author}{S.~Venkataraman}, \bibinfo{author}{D.~Liu},
  \bibinfo{author}{J.~Freeman}, \bibinfo{author}{D.~Tsai},
  \bibinfo{author}{M.~Amde}, \bibinfo{author}{S.~Owen}, et~al.,
\newblock \bibinfo{title}{Mllib: Machine learning in apache spark},
\newblock \bibinfo{journal}{The Journal of Machine Learning Research}
  \bibinfo{volume}{17} (\bibinfo{year}{2016}) \bibinfo{pages}{1235--1241}.
\bibitem[{Nitsche and Halbritter(2019)}]{nitsche2019development}
\bibinfo{author}{M.~Nitsche}, \bibinfo{author}{S.~Halbritter},
\newblock \bibinfo{title}{Development of an end-to-end deep learning pipeline},
\newblock \bibinfo{journal}{Hochschule f{\"u}r Angewandte Wissenschaften
  Hamburg}  (\bibinfo{year}{2019}).
\bibitem[{Shaikh et~al.(2017)Shaikh, Vishwakarma, Mehta, Varshney, Ramamurthy,
  and Wei}]{shaikh2017end}
\bibinfo{author}{S.~Shaikh}, \bibinfo{author}{H.~Vishwakarma},
  \bibinfo{author}{S.~Mehta}, \bibinfo{author}{K.~R. Varshney},
  \bibinfo{author}{K.~N. Ramamurthy}, \bibinfo{author}{D.~Wei},
\newblock \bibinfo{title}{An end-to-end machine learning pipeline that ensures
  fairness policies},
\newblock \bibinfo{journal}{arXiv preprint arXiv:1710.06876}
  (\bibinfo{year}{2017}).
\bibitem[{Boovaraghavan et~al.(2021)Boovaraghavan, Maravi, Mallela, and
  Agarwal}]{boovaraghavan2021mliot}
\bibinfo{author}{S.~Boovaraghavan}, \bibinfo{author}{A.~Maravi},
  \bibinfo{author}{P.~Mallela}, \bibinfo{author}{Y.~Agarwal},
\newblock \bibinfo{title}{Mliot: An end-to-end machine learning system for the
  internet-of-things},
\newblock in: \bibinfo{booktitle}{Proceedings of the International Conference
  on Internet-of-Things Design and Implementation}, \bibinfo{year}{2021}, pp.
  \bibinfo{pages}{169--181}.
\bibitem[{Molinara et~al.(2020)Molinara, Ferdinandi, Cerro, Ferrigno, and
  Massera}]{molinara2020end}
\bibinfo{author}{M.~Molinara}, \bibinfo{author}{M.~Ferdinandi},
  \bibinfo{author}{G.~Cerro}, \bibinfo{author}{L.~Ferrigno},
  \bibinfo{author}{E.~Massera},
\newblock \bibinfo{title}{An end to end indoor air monitoring system based on
  machine learning and sensiplus platform},
\newblock \bibinfo{journal}{IEEE Access} \bibinfo{volume}{8}
  (\bibinfo{year}{2020}) \bibinfo{pages}{72204--72215}.
\bibitem[{Vinzamuri et~al.(2020)Vinzamuri, Khabiri, Bhamidipaty, Mckim, and
  Gandhi}]{vinzamuri2020end}
\bibinfo{author}{B.~Vinzamuri}, \bibinfo{author}{E.~Khabiri},
  \bibinfo{author}{A.~Bhamidipaty}, \bibinfo{author}{G.~Mckim},
  \bibinfo{author}{B.~Gandhi},
\newblock \bibinfo{title}{An end-to-end context aware anomaly detection
  system},
\newblock in: \bibinfo{booktitle}{2020 IEEE International Conference on Big
  Data (Big Data)}, \bibinfo{organization}{IEEE}, \bibinfo{year}{2020}, pp.
  \bibinfo{pages}{1689--1698}.
\bibitem[{Li et~al.(2020)Li, Zha, Venugopal, Zou, and Hu}]{li2020pyodds}
\bibinfo{author}{Y.~Li}, \bibinfo{author}{D.~Zha},
  \bibinfo{author}{P.~Venugopal}, \bibinfo{author}{N.~Zou},
  \bibinfo{author}{X.~Hu},
\newblock \bibinfo{title}{Pyodds: An end-to-end outlier detection system with
  automated machine learning},
\newblock in: \bibinfo{booktitle}{Companion Proceedings of the Web Conference
  2020}, \bibinfo{year}{2020}, pp. \bibinfo{pages}{153--157}.
\bibitem[{Fezari and Al~Dahoud(2018)}]{fezari2018integrated}
\bibinfo{author}{M.~Fezari}, \bibinfo{author}{A.~Al~Dahoud},
\newblock \bibinfo{title}{Integrated development environment “ide” for
  arduino},
\newblock \bibinfo{journal}{WSN applications}  (\bibinfo{year}{2018})
  \bibinfo{pages}{1--12}.
\bibitem[{Hiertz et~al.(2010)Hiertz, Denteneer, Stibor, Zang, Costa, and
  Walke}]{hiertz2010ieee}
\bibinfo{author}{G.~R. Hiertz}, \bibinfo{author}{D.~Denteneer},
  \bibinfo{author}{L.~Stibor}, \bibinfo{author}{Y.~Zang},
  \bibinfo{author}{X.~P. Costa}, \bibinfo{author}{B.~Walke},
\newblock \bibinfo{title}{The ieee 802.11 universe},
\newblock \bibinfo{journal}{IEEE Communications Magazine} \bibinfo{volume}{48}
  (\bibinfo{year}{2010}) \bibinfo{pages}{62--70}.
\bibitem[{Chang(2014)}]{chang2014bluetooth}
\bibinfo{author}{K.-H. Chang},
\newblock \bibinfo{title}{Bluetooth: a viable solution for iot?[industry
  perspectives]},
\newblock \bibinfo{journal}{IEEE Wireless Communications} \bibinfo{volume}{21}
  (\bibinfo{year}{2014}) \bibinfo{pages}{6--7}.
\bibitem[{Heydon and Hunn(2012)}]{heydon2012bluetooth}
\bibinfo{author}{R.~Heydon}, \bibinfo{author}{N.~Hunn},
\newblock \bibinfo{title}{Bluetooth low energy},
\newblock \bibinfo{journal}{CSR Presentation, Bluetooth SIG https://www.
  bluetooth. org/DocMan/handlers/DownloadDoc. ashx}  (\bibinfo{year}{2012}).
\bibitem[{Ergen(2004)}]{ergen2004zigbee}
\bibinfo{author}{S.~C. Ergen},
\newblock \bibinfo{title}{Zigbee/ieee 802.15. 4 summary},
\newblock \bibinfo{journal}{UC Berkeley, September} \bibinfo{volume}{10}
  (\bibinfo{year}{2004}) \bibinfo{pages}{11}.
\bibitem[{Song et~al.(2008)Song, Han, Mok, Chen, Lucas, Nixon, and
  Pratt}]{song2008wirelesshart}
\bibinfo{author}{J.~Song}, \bibinfo{author}{S.~Han}, \bibinfo{author}{A.~Mok},
  \bibinfo{author}{D.~Chen}, \bibinfo{author}{M.~Lucas},
  \bibinfo{author}{M.~Nixon}, \bibinfo{author}{W.~Pratt},
\newblock \bibinfo{title}{Wirelesshart: Applying wireless technology in
  real-time industrial process control},
\newblock in: \bibinfo{booktitle}{2008 IEEE Real-Time and Embedded Technology
  and Applications Symposium}, \bibinfo{organization}{IEEE},
  \bibinfo{year}{2008}, pp. \bibinfo{pages}{377--386}.
\bibitem[{Warden and Situnayake(2019)}]{warden2019tinyml}
\bibinfo{author}{P.~Warden}, \bibinfo{author}{D.~Situnayake},
  \bibinfo{title}{Tinyml: Machine learning with tensorflow lite on arduino and
  ultra-low-power microcontrollers}, \bibinfo{publisher}{" O'Reilly Media,
  Inc."}, \bibinfo{year}{2019}.
\bibitem[{Node-{RED}(2021)}]{Node-RED}
\bibinfo{author}{Node-{RED}}, \bibinfo{title}{Node-{RED}},
  \bibinfo{year}{2021}. \URLprefix \url{https://nodered.org/}.
\bibitem[{Tilkov and Vinoski(2010)}]{tilkov2010node}
\bibinfo{author}{S.~Tilkov}, \bibinfo{author}{S.~Vinoski},
\newblock \bibinfo{title}{Node. js: Using javascript to build high-performance
  network programs},
\newblock \bibinfo{journal}{IEEE Internet Computing} \bibinfo{volume}{14}
  (\bibinfo{year}{2010}) \bibinfo{pages}{80--83}.
\bibitem[{Arduino(2021)}]{arduinoBLE}
\bibinfo{author}{Arduino}, \bibinfo{title}{Arduino {Nano} 33 {BLE} {Sense}
  {\textbar} {Arduino} {Official} {Store}}, \bibinfo{year}{2021}. \URLprefix
  \url{https://store.arduino.cc/arduino-nano-33-ble-sense}.
\bibitem[{Semiconductor(2021)}]{nrf52840}
\bibinfo{author}{N.~Semiconductor}, \bibinfo{title}{{nRF52840} - {Nordic}
  {Semiconductor}}, \bibinfo{year}{2021}. \URLprefix
  \url{nordicsemi.com/Products/nRF52840}.
\bibitem[{IEEE(2019)}]{8766229}
\bibinfo{author}{IEEE},
\newblock \bibinfo{title}{Ieee standard for floating-point arithmetic},
\newblock \bibinfo{journal}{IEEE Std 754-2019 (Revision of IEEE 754-2008)}
  (\bibinfo{year}{2019}) \bibinfo{pages}{1--84}.
  \DOIprefix\doi{10.1109/IEEESTD.2019.8766229}.
\bibitem[{Yao and Warren(2005)}]{yao2005applying}
\bibinfo{author}{J.~Yao}, \bibinfo{author}{S.~Warren},
\newblock \bibinfo{title}{Applying the iso/ieee 11073 standards to wearable
  home health monitoring systems},
\newblock \bibinfo{journal}{Journal of clinical monitoring and computing}
  \bibinfo{volume}{19} (\bibinfo{year}{2005}) \bibinfo{pages}{427--436}.
\bibitem[{Badamasi(2014)}]{badamasi2014working}
\bibinfo{author}{Y.~A. Badamasi},
\newblock \bibinfo{title}{The working principle of an arduino},
\newblock in: \bibinfo{booktitle}{2014 11th international conference on
  electronics, computer and computation (ICECCO)},
  \bibinfo{organization}{IEEE}, \bibinfo{year}{2014}, pp.
  \bibinfo{pages}{1--4}.
\bibitem[{Due and Core(2017)}]{due2017arduino}
\bibinfo{author}{A.~Due}, \bibinfo{author}{A.~Core},
\newblock \bibinfo{title}{Arduino due},
\newblock \bibinfo{journal}{Retrieved} \bibinfo{volume}{9}
  (\bibinfo{year}{2017}) \bibinfo{pages}{2019}.
\bibitem[{Grove(2021)}]{grove}
\bibinfo{author}{Grove}, \bibinfo{title}{Sensors - {Seeed} {Studio}
  {Electronics}}, \bibinfo{year}{2021}. \URLprefix
  \url{https://www.seeedstudio.com/category}.
\bibitem[{Quick and Choo(2014)}]{quick2014google}
\bibinfo{author}{D.~Quick}, \bibinfo{author}{K.-K.~R. Choo},
\newblock \bibinfo{title}{Google drive: Forensic analysis of data remnants},
\newblock \bibinfo{journal}{Journal of Network and Computer Applications}
  \bibinfo{volume}{40} (\bibinfo{year}{2014}) \bibinfo{pages}{179--193}.
\bibitem[{Carneiro et~al.(2018)Carneiro, Da~N{\'o}brega, Nepomuceno, Bian,
  De~Albuquerque, and Reboucas~Filho}]{carneiro2018performance}
\bibinfo{author}{T.~Carneiro}, \bibinfo{author}{R.~V.~M. Da~N{\'o}brega},
  \bibinfo{author}{T.~Nepomuceno}, \bibinfo{author}{G.-B. Bian},
  \bibinfo{author}{V.~H.~C. De~Albuquerque}, \bibinfo{author}{P.~P.
  Reboucas~Filho},
\newblock \bibinfo{title}{Performance analysis of google colaboratory as a tool
  for accelerating deep learning applications},
\newblock \bibinfo{journal}{IEEE Access} \bibinfo{volume}{6}
  (\bibinfo{year}{2018}) \bibinfo{pages}{61677--61685}.
\bibitem[{Giraldo et~al.(2018)Giraldo, Urbina, Cardenas, Valente, Faisal,
  Ruths, Tippenhauer, Sandberg, and Candell}]{giraldo2018survey}
\bibinfo{author}{J.~Giraldo}, \bibinfo{author}{D.~Urbina},
  \bibinfo{author}{A.~Cardenas}, \bibinfo{author}{J.~Valente},
  \bibinfo{author}{M.~Faisal}, \bibinfo{author}{J.~Ruths},
  \bibinfo{author}{N.~O. Tippenhauer}, \bibinfo{author}{H.~Sandberg},
  \bibinfo{author}{R.~Candell},
\newblock \bibinfo{title}{A survey of physics-based attack detection in
  cyber-physical systems},
\newblock \bibinfo{journal}{ACM Computing Surveys (CSUR)} \bibinfo{volume}{51}
  (\bibinfo{year}{2018}) \bibinfo{pages}{1--36}.
\bibitem[{Barrett and Pack(2012)}]{barrett2012atmel}
\bibinfo{author}{S.~F. Barrett}, \bibinfo{author}{D.~J. Pack},
\newblock \bibinfo{title}{Atmel avr microcontroller primer: Programming and
  interfacing},
\newblock \bibinfo{journal}{Synthesis Lectures on Digital Circuits and Systems}
  \bibinfo{volume}{7} (\bibinfo{year}{2012}) \bibinfo{pages}{1--244}.
\bibitem[{Bai(2015)}]{bai2015practical}
\bibinfo{author}{Y.~Bai}, \bibinfo{title}{Practical microcontroller engineering
  with ARM technology}, \bibinfo{publisher}{John Wiley \& Sons},
  \bibinfo{year}{2015}.
\bibitem[{Hasan et~al.(2019)Hasan, Islam, Zarif, and Hashem}]{hasan2019attack}
\bibinfo{author}{M.~Hasan}, \bibinfo{author}{M.~M. Islam},
  \bibinfo{author}{M.~I.~I. Zarif}, \bibinfo{author}{M.~Hashem},
\newblock \bibinfo{title}{Attack and anomaly detection in iot sensors in iot
  sites using machine learning approaches},
\newblock \bibinfo{journal}{Internet of Things} \bibinfo{volume}{7}
  (\bibinfo{year}{2019}) \bibinfo{pages}{100059}.
\bibitem[{Davis and Goadrich(2006)}]{davis2006relationship}
\bibinfo{author}{J.~Davis}, \bibinfo{author}{M.~Goadrich},
\newblock \bibinfo{title}{The relationship between precision-recall and roc
  curves},
\newblock in: \bibinfo{booktitle}{Proceedings of the 23rd international
  conference on Machine learning}, \bibinfo{year}{2006}, pp.
  \bibinfo{pages}{233--240}.
\bibitem[{Gulli and Pal(2017)}]{gulli2017deep}
\bibinfo{author}{A.~Gulli}, \bibinfo{author}{S.~Pal}, \bibinfo{title}{Deep
  learning with Keras}, \bibinfo{publisher}{Packt Publishing Ltd},
  \bibinfo{year}{2017}.
\bibitem[{Google(2021)}]{GoogleTFMicro2021}
\bibinfo{author}{Google}, \bibinfo{title}{{TensorFlow Lite Micro}},
  \bibinfo{year}{2021}. \URLprefix
  \url{https://www.tensorflow.org/lite/microcontrollers}.
\bibitem[{TensorFlow(2021)}]{tensorflow_converter}
\bibinfo{author}{TensorFlow}, \bibinfo{title}{{TensorFlow} {Lite} converter},
  \bibinfo{year}{2021}. \URLprefix
  \url{https://www.tensorflow.org/lite/convert}.
\bibitem[{Banbury et~al.(2020)Banbury, Reddi, Lam, Fu, Fazel, Holleman, Huang,
  Hurtado, Kanter, Lokhmotov et~al.}]{banbury2020benchmarking}
\bibinfo{author}{C.~R. Banbury}, \bibinfo{author}{V.~J. Reddi},
  \bibinfo{author}{M.~Lam}, \bibinfo{author}{W.~Fu},
  \bibinfo{author}{A.~Fazel}, \bibinfo{author}{J.~Holleman},
  \bibinfo{author}{X.~Huang}, \bibinfo{author}{R.~Hurtado},
  \bibinfo{author}{D.~Kanter}, \bibinfo{author}{A.~Lokhmotov}, et~al.,
\newblock \bibinfo{title}{Benchmarking tinyml systems: Challenges and
  direction},
\newblock \bibinfo{journal}{arXiv preprint arXiv:2003.04821}
  (\bibinfo{year}{2020}).
\bibitem[{Python(ckle)}]{pickle}
\bibinfo{author}{Python}, \bibinfo{title}{pickle — {Python} object
  serialization — {Python} 3.9.6 documentation}, \bibinfo{year}{pickle}.
  \URLprefix \url{https://docs.python.org/3/library/pickle.html}.

\end{thebibliography}
\end{document}